\documentclass[structabstract]{aa}  
\usepackage{float}
\usepackage{graphicx}
\usepackage{longtable}
\usepackage{txfonts}
\begin{document}
\title{A VLT/FLAMES survey for massive binaries in Westerlund 1 V. the 
X-ray selected blue stragglers Wd1-27 and -30a
\thanks{Based on observations made at the European Southern Observatory, 
Paranal, Chile under programs ESO 081.D-0324, 383.D-0633,  
087.D-0440, 091.D-0179, and 097.D-0367} 
}
\author{J.~S.~Clark\inst{1}
\and F.~Najarro\inst{2}
\and I.~Negueruela\inst{3}
\and B.~W.~Ritchie\inst{1}
\and C.~Gonz\'{a}lez-Fern\'{a}ndez\inst{4}
\and M.~E.~Lohr\inst{1}}
\institute{
$^1$School of Physical Science, The Open 
University, Walton Hall, Milton Keynes, MK7 6AA, United Kingdom\\
$^2$Departamento de Astrof\'{\i}sica, Centro de Astrobiolog\'{\i}a, 
(CSIC-INTA), Ctra. Torrej\'on a Ajalvir, km 4,  28850 Torrej\'on de Ardoz, 
Madrid, Spain\\
$^3$ Departamento de F\'{\i}sica Aplicada, Facultad de
Ciencias, Universidad de Alicante, Carretera San Vicente del Raspeig s/n,
E03690, San Vicente del Raspeig, Spain\\
$^4$Institute of Astronomy, University of Cambridge, Madingley Road,
Cambridge CB3 0HA, United Kingdom}

   \abstract{Recent observational studies indicate that a large number of OB stars are found within 
binary systems which may be expected to interact during their lifetimes. Significant mass transfer or indeed merger of both components is  expected to modify evolutionary pathways, facilitating 
the production of exceptionally massive stars which will present as blue stragglers. Identification and characterisation of 
such objects is crucial if the efficiency of mass transfer is to be established; a critical parameter in determining the 
 outcomes of binary evolutionary channels.  }
{The young and coeval massive cluster Westerlund 1 hosts a rich population of X-ray bright OB and Wolf-Rayet stars where the
 emission is attributed to shocks in the wind collision zones of massive binaries. Motivated by this, we instigated a study of the  extremely X-ray luminous O supergiants Wd1-27 and -30a. }
{We subjected a multi-wavelength and -epoch photometric and spectroscopic dataset to quantitative non-LTE model atmosphere and 
time-series analysis in order to determine fundamental stellar parameters and search for evidence of binarity. A detailed examination of the second Gaia data release was undertaken to establish cluster membership.}
{Both stars were found to be early/mid-O hypergiants with luminosities, temperatures and masses significantly in excess of other 
early stars within Wd1, hence qualifying as massive blue stragglers. The binary nature of Wd1-27 remains uncertain but the detection of radial velocity changes and the X-ray properties of Wd1-30a suggest that it is a binary with an orbital period $\leq10$ days. Analysis of Gaia proper motion and parallactic data indicates that both stars are cluster members; we also provide a membership list for Wd1 based on this analysis.}{The presence of hypergiants of spectral types O to M within Wd1 cannot be understood solely via single-star evolution. We suppose that the early-B  and mid-O hypergiants formed via binary-induced mass-stripping of the primary and mass-transfer to the secondary, respectively. This implies that for a subset of objects  massive star-formation may be regarded as a two-stage process, with binary-driven mass-transfer or merger yielding stars with masses significantly in excess of their initial `birth' mass.   
}


\maketitle

\section{Introduction}

Given the importance of radiative and mechanical feedback from massive stars to galactic 
evolution,  and their role as the progenitors of  electromagnetic and, ultimately,  
gravitational wave transients, current  uncertainties regarding  the physics of many stages of 
their lifecycles is a serious concern. A particular issue is the mechanism by which they form. 
Reviews of this process by Zinnecker \& Yorke (\cite{zinnecker}) and Krumholz (\cite{krumholz}) 
suggest two `families' of models - accretion or merger. The first comprises both the 
fragmentation and subsequent monolithic collapse of a molecular cloud - essentially a 
scaled-up version of low-mass star-formation - and the competitive accretion scenario of 
Bonnell et al. (\cite{bonnell01}), which occurs in a (proto-)clustered environment.  The second 
scenario envisages the formation of very massive objects by the merger of lower-mass 
(proto-)stars (e.g. Bonnell et al. \cite{bonnell98}); under such a scenario massive star 
formation becomes a multi-stage process.

However, as with competitive accretion, collisional merger requires (exceptionally) dense 
stellar environments to be viable, and it is not clear that 
even clusters as extreme as the Arches and R136 supply the required conditions (Krumholz 
\cite{krumholz}). Moreover recent observational findings  challenge the assertion 
that massive stars form exclusively in highly clustered  environments, with  Rosslowe \& 
Crowther (\cite{rosslowe}) reporting that only $\sim25$\% of galactic Wolf-Rayets (WRs) are associated with 
young massive clusters (YMCs). Likewise Wright et al. (\cite{wright16}) demonstrate that the Cyg 
OB2 association - and the high-mass stars that formed within it - did not originate in high-density 
clusters that subsequently dissolved into the wider environment, instead  being born in  the current
dispersed  configuration.

Nevertheless, there has been considerable recent interest in a variant of the merger scenario, 
in which binary interaction leads  to the rejuvenation of the binary product via mass transfer and/or merger
(e.g.  van Bever \& Vanbeveren \cite{vbvb})\footnote{See 
also Banerjee et al. (\cite{banerjeea}, \cite{banerjeeb}), who  suggest very massive stars 
may form in  dense environments via dynamically-induced mergers of massive binaries, rather 
than being driven by stellar evolution.}. Such attention has been driven by two related 
observational assertions; that the core of R136 may contain a handful of stars with masses 
exceeding the canonical upper mass limit of $\sim150M_{\odot}$ (Crowther et al. 
\cite{crowther10}, Oey \& Clarke \cite{oey}) and that the most massive stars in e.g. the 
Arches and Quintuplet YMCs appear younger than lower-mass cluster members (e.g. Martins et al.  
\cite{martins08}, Liermann et al. \cite{liermann12}; but see Sect. 5.2 for discussion of recent 
countervailing analyses). Schneider et al. 
(\cite{schneider14}, \cite{schneider15}; see also de Mink et al. \cite{demink}) were able to 
replicate these findings under the  assumption of high binary fractions for both clusters, 
with the most luminous stars being post-binary interaction systems; essentially the
high-mass analogues of the classical `blue  stragglers' seen in globular clusters (e.g. Sandage \cite{sandage}).

That binary-driven mass transfer or merger should lead to rejuvenation is an uncontroversial 
statement, with both Algols and W Serpentis stars serving as exemplars (e.g. Tarasov 
\cite{tarasov}). A more massive analogue would be the interacting binary RY Scuti, where the primary is less 
massive than the secondary, which is  currently veiled by an accretion disc (e.g. $7.1\pm1.2M_{\odot}$ and 
$30.0\pm2.1M_{\odot}$; Grundstrom et al. \cite{grundstrom}). An example of a massive post-interaction 
system is NGC346-13, where the 
more evolved early-B giant is less massive than its late-O dwarf companion 
($11.9\pm0.6M_{\odot}$ and $19.1\pm1.0M_{\odot}$; Ritchie et al. \cite{ritchie12}). Unfortunately
RY Scuti is not associated with a cluster, while NGC346 has experienced a  complex star 
formation history over at least 6Myr (Cignoni et al. \cite{cignoni}). As a consequence it 
is difficult to reconstruct the mass-transfer history of either system to determine the quantity of mass 
transferred to the secondary (and that lost to the system) and hence whether they represent  `bona fide' 
massive blue stragglers; one is instead forced to rely on theoretical predictions 
(e.g. Petrovic et al. \cite{petrovic}) which are inevitably subject to uncertainties in the input physics.

\subsection{The YMC Westerlund 1}

Given its comparative youth and exceptional integrated mass, the galactic cluster Westerlund 1 (Wd1) would 
appear to be an ideal  laboratory to search for the products of  binary interaction (Clark et al. 
\cite{clark05}). It appears to be highly co-eval (Negueruela et al. \cite{iggy10}, Kudryavtseva et 
al. \cite{ku}) and, as a result, is characterised by a  remarkably  homogeneous population of early 
supergiants, extending smoothly in  spectral morphology from $\sim$O9.5-B4 Ia (Negueruela et al. 
\cite{iggy10}, Clark et al. \cite{clark15}).  Intensive multiwavelength and multi-epoch observational campaigns 
(e.g. Clark et al. \cite{clark08},  \cite{clark11};  Ritchie et al. \cite{ritchie09},  \cite{ritchie10}, 
in prep.) have revealed a rich  binary population comprising pre-interaction (e.g. Wd1-43a; Ritchie et al. 
\cite{ritchie11}),  interacting (Wd1-9; Clark et al. \cite{clark13},  Fenech et al.  \cite{fenech17}) and 
post-interaction binaries (Wd1-13 and -239; Ritchie et al. \cite{ritchie10}, Clark et al. \cite{clark11}). 

Historically X-ray data have proved extremely useful in identifying massive interactive binaries.
Single OB stars are expected to emit X-rays of luminosity $L_{\rm X} \sim 10^{-7}L_{\rm bol}$ 
via shocks embedded within their stellar winds (Long \& White \cite{long}, Lucy \& White \cite{lucy},
Seward \& Chlebowski \cite{seward}, Berghoefer et al. \cite{berg}). Detailed analyses reveal 
 thermal spectra for such sources, with a  characteristic energy of $kT=0.6$keV (Gayley \& Owocki 
\cite{gayley}, Feldmeier et al. \cite{feldmeier}). Stars demonstrating greater luminosities 
and/or harder X-ray spectra are typically assumed to be massive binaries, with the excess (hard) 
emission arising in shocks generated by their colliding winds.

For stellar luminosities of $L_{\rm bol}\sim6\times10^5L_{\odot}$ we 
would expect $L_{\rm X} \sim 2 \times 10^{32}$erg s$^{-1}$ for {\em single} late-O/early B supergiants within Wd1 (Negueruela 
et al. \cite{iggy10}).
However a number of  supergiants are found to be more luminous than this, with Wd1-27 and -53 approaching 
and Wd1-30a and -36 in excess of $L_{\rm X} \sim 10^{33}$erg s$^{-1}$ (Clark et al. \cite{clark08}). 
Of these the X-ray spectra of Wd1-27, -36 and -53 are comparatively soft ($kT{\sim}0.5-0.7$keV) and hence 
consistent with emission from a single star while, 
with  $kT=1.3\pm0.1$keV, the emission from Wd1-30 is substantially harder than this expectation.
Photometric monitoring  of Wd1-36  and -53 reveals periodic modulation with periods of 3.18d 
and 1.3d respectively (Bonanos et al. \cite{bonanos}). Wd1-36 is clearly an eclipsing system, while the 
lightcurve of  Wd-53 likely results from ellipsoidal  modulation; hence both appear compelling binary 
candidates. However to date there is no corroborative evidence for binarity for Wd1-27 and -30a and in this
paper we investigate their nature with a multi-epoch optical and near-IR spectroscopic dataset.

\section{Data acquisition and  reduction}

\begin{figure*}
\includegraphics[width=12cm,angle=-90]{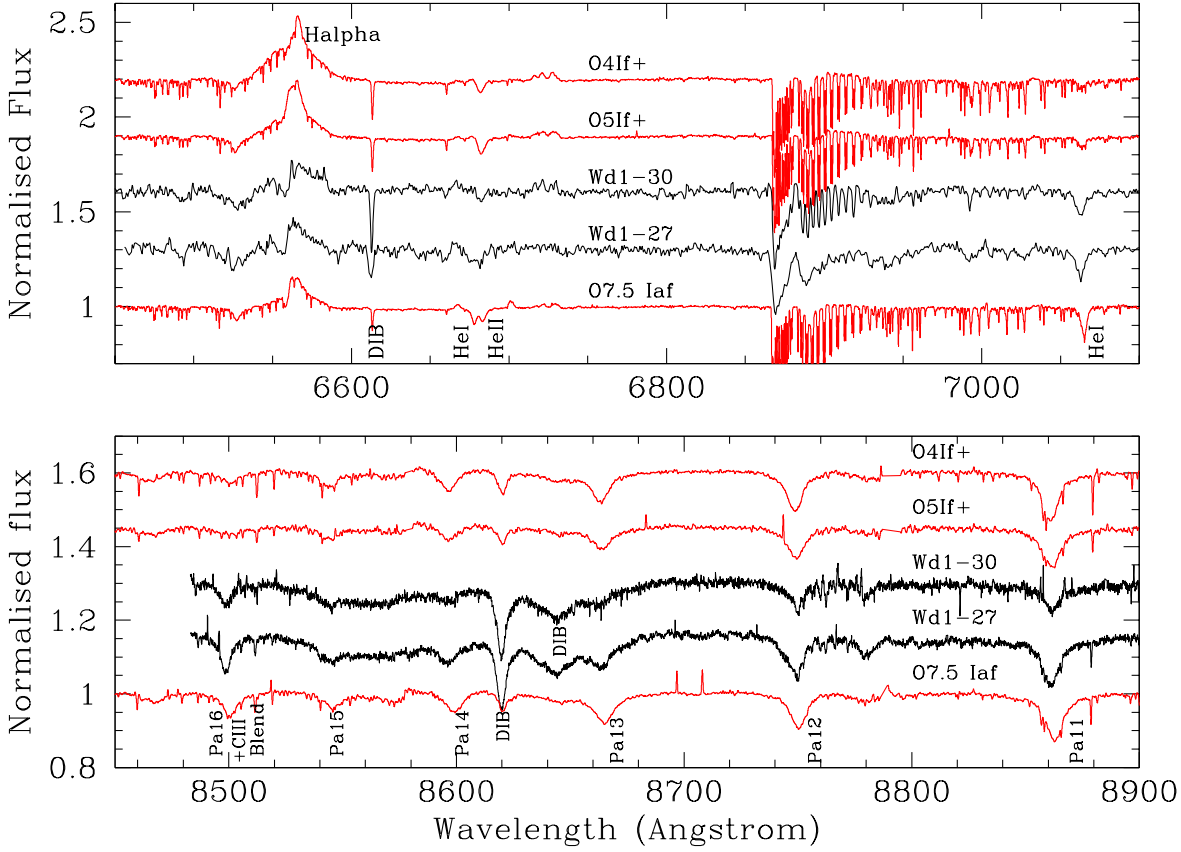}
\caption{$R-$ and $I-$band spectra of Wd1-27 and -30a (black) with prominent
transitions indicated (note the difference in spectral resolutions between 
the two observations). Template spectra of  HD15570 (O4 If$^+$), HD14947 (O5
If$^+$) and  HD192639 (O7.5 Iaf) are shown for comparison (red). }
\end{figure*}

\begin{figure*}
\includegraphics[width=17cm,angle=0]{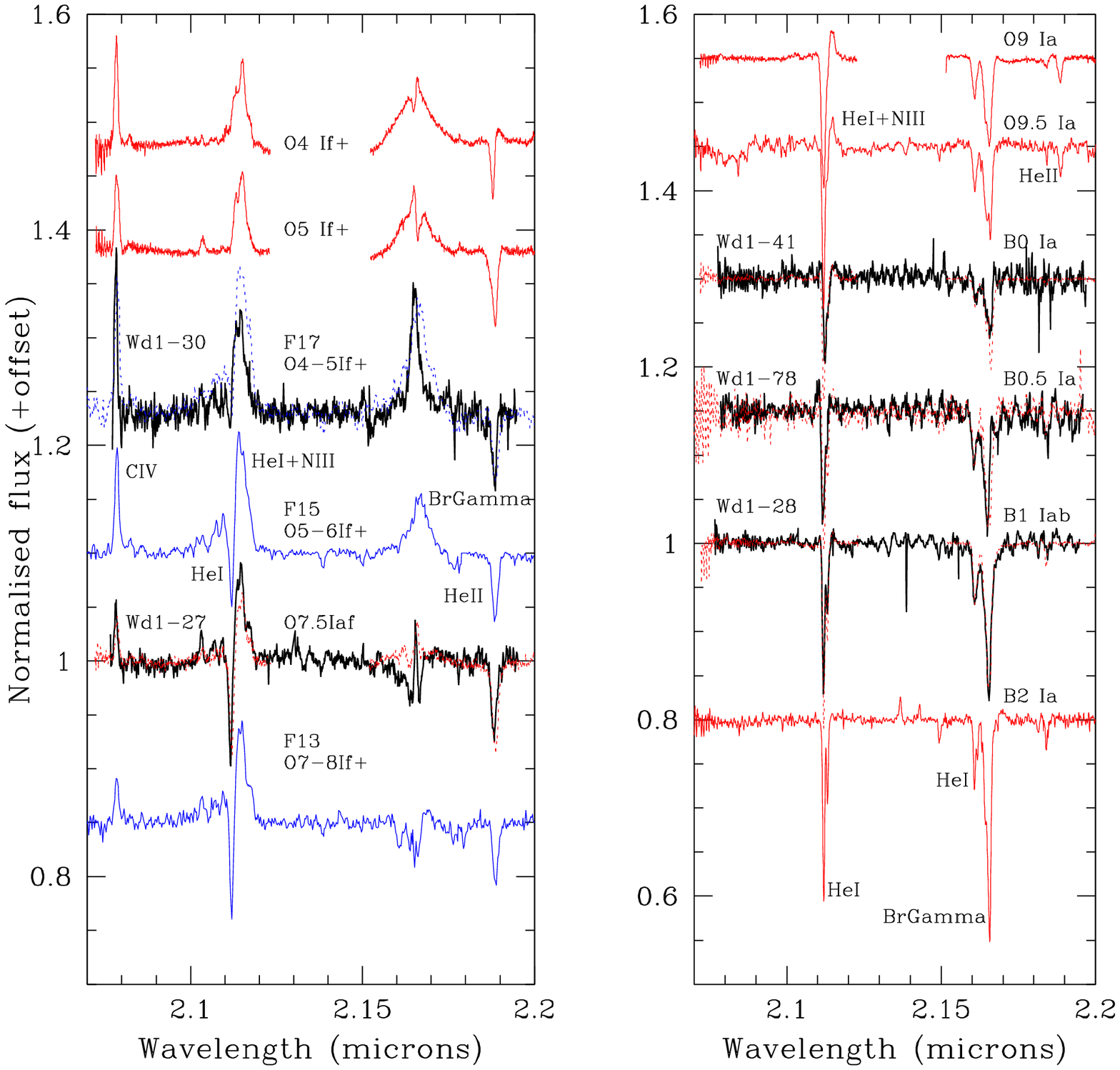}
\caption{Left panel: K-band spectra of Wd1-27 and -30a (black) with
prominent transitions indicated. Template spectra of  HD15570 (O4 If$^+$),
HD14947 (O5 If$^+$) and  HD192639 (O7.5 Iaf) shown for comparison (red; data
from Hanson et al. \cite{hanson05}). Additional comparators from the
Arches cluster are plotted in blue (data from Clark et al. \cite{clark18a}).
Right panel: spectra of
`representative' early-B cluster supergiants Wd1-28, -41 and -78 shown for
comparison (black) with additional template spectra (red; Hanson et al.
\cite{hanson05}) of HD30614 (O9 Ia), HD154368 (O9.5 Iab), HD37128 (B0 Ia),
HD115842 (B0.5 Ia) and  HD13854 (B1 Iab).}
\end{figure*}

\subsection{Spectroscopy}

A single spectrum of Wd1-27 was obtained on 2006 February 17 with the 
NTT/EMMI with grism \#6 covering  the range 5800-8650{\AA} at
a resolution R$\sim$1500; reduction details may be found in Negueruela et 
al. (\cite{iggy10}).  This is presented in Fig. 1 and encompasses the 
prime mass-loss diagnostic  H$\alpha$, the higher Paschen series and a 
selection of He\,{\sc i} and He\,{\sc ii} photospheric features.

Subsequently one and three epochs of $I-$band observations were made during 2011 and  2013 
April-September, respectively, with VLT/FLAMES. We  utilised the  GIRAFFE 
spectrograph in MEDUSA mode with setup HR21 to cover the 8484-9001{\AA} 
range with resolution R$\sim$16,200; full details
of data acquisition and reduction are given in Ritchie et al. 
(\cite{ritchie09}). The resultant spectra encompass the Paschen series 
photospheric lines  and were obtained with the intention of searching for 
radial velocity (RV)  variability. However the presence of temperature 
sensitive He\,{\sc i} 
photospheric features also permits spectral classification utilising such 
data (e.g. Clark et al. \cite{clark05}).

A further three spectra were obtained 
between 2016 May-June with VLT/UVES (Dekker et al. \cite{dekker}).
In each observation run, two 1482 s exposures were taken sequentially, and 
the blue and red arms
were operated in parallel using the dichroic beam splitter, resulting in
usable wavelength coverage over the ranges $\sim5695 - 7530${\AA} and
$\sim7660 - 9460${\AA}.  The 0.7" slit was used, giving a resolving
power R$\sim$60,000.  Basic reductions (bias subtraction, interorder 
background subtraction, flat-field correction, echelle order extraction, 
sky subtraction, rebinning to wavelength scale and order merging) were 
carried out using the ESO UVES pipeline software (version 5.7.0) running 
under Gasgano.  A custom IDL code was then used to identify and remove bad 
lines by comparing the two exposures for each epoch, before summing the 
cleaned spectra.

Turning to  Wd1-30a and  eleven epochs of $I-$band spectroscopy,  also 
utilising the GIRAFFE+HR21 configuration of 
VLT/FLAMES, were made between 2008 June - 2009 August, with a further five 
epochs between 2013 April - September. These were supplemented by a single 
observation with VLT/FLAMES with the low-resolution grating LR6,
yielding  an $R-$band spectrum with a wavelength range of 6438-7184{\AA} at a 
resolution of R$\sim8500$ that was designed to sample H$\alpha$. Examples 
of the former and  latter spectra are presented in Fig. 1.

Finally, $K-$band spectra  of Wd1-27, 30a and other evolved stars 
within Wd1 were obtained with   VLT/ISAAC between 2004 August 
1-3 under programme 073.D-0837  and were extracted from the ESO archive. 
Observations were made in the short wavelength mode and employed the medium 
resolution grating  with a central wavelength of 2.13825$\mu$m and 
0.6" slit, yielding a resolution of R$\sim4400$ between 
$\sim2.077-2.199\mu$m. The spectra were reduced and extracted using the 
ISAAC pipeline provided by ESO. Due to the high stellar density, often two 
or more objects fall into the slit, and particular care was taken in 
extracting all these secondary
spectra, along with the identification of the corresponding sources. The
wavelength calibration was refined using the abundant telluric features
present in the NIR, and a transmission curve of the atmosphere+telescope
system built using observations of telluric standards. These comprised 
both early-type and solar-like stars; for the latter the subtle 
mismatches between the stellar absorption lines in each adopted model and 
the observed standard were corrected by fitting these lines using a family 
of gaussians. Observations of targets were corrected utilising the 
resultant transmission curves before continuum nomalisation was 
undertaken.  A subset of the resulting spectra  are presented in 
Fig. 2, comprising Wd1-27, -30a and a representative group of supergiants
 in order to place these into context.

\subsection{Photometry}

 Optical and near-IR photometry derive  from  Clark et al. 
(\cite{clark05}), Negueruela et al. (\cite{iggy10}) and Crowther et  al. 
(\cite{crowther06}) and are summarised in Table 1. Due to crowding and 
saturation no mid-IR fluxes may be determined for either source.  Wd1-30a has 
a 3mm flux of 0.17$\pm0.06$mJy (Fenech et al. \cite{fenech18}), while Wd1-27 is a
non-detection with a $3\sigma$ upper limit of 0.13mJ; neither star is 
detected at
radio wavelengths (3.6cm and longer) with $3\sigma$ upper limits of 
0.17mJy (Dougherty et al. \cite{dougherty}).

 Finally, neither star is reported  to be a short- or long-term photometric 
variable (timescale of $\sim$days and $\sim$ years respectively; Bonanos et al. \cite{bonanos},
Clark et al. \cite{clark10}).

\begin{table}
\caption{Optical and near-IR photometry}
\begin{center}
\begin{tabular}{lccccccc}
\hline
\hline
Star   & B & V & R & I & J & H & K \\
\hline
Wd1-27  & 21.5  & 17.94  & 15.35  & 12.80  & 9.98  & 8.92  & 8.49  \\
Wd1-30a & 22.4  & 18.45  & 15.80  & 13.20  & 10.47  & 9.42  & 9.05  \\
\hline
\end{tabular}
\end{center}
{Errors are $\sim0.1$m in B-band,  $\sim0.02$m in V-, R- and I-bands and 
$\sim0.05$m in J-, H- and K-bands.} 
\end{table}

\section{Observational properties and cluster membership}

\begin{figure*}
\includegraphics[width=14cm,angle=0]{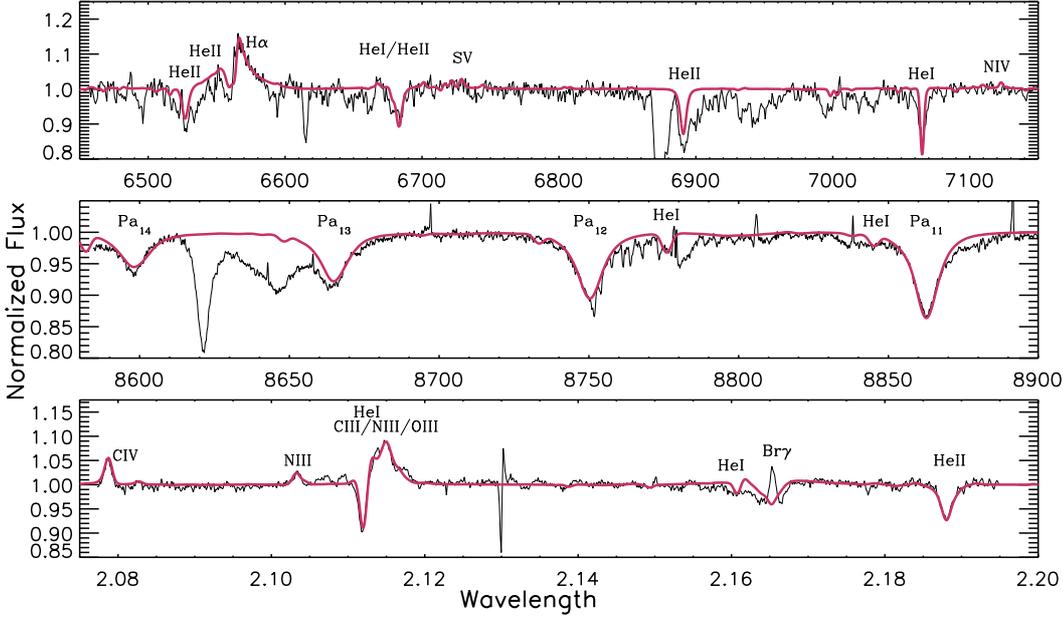}
\caption{Comparison of the synthetic spectrum of Wd1-27 derived from model-atmosphere analysis (red dashed line) to observational 
data (black solid line). See Sect. 4 for further details. The units of wavelength for the top and middle panels are Angstroms 
and the bottom panel microns.}
\end{figure*}

\begin{figure*}
\includegraphics[width=14cm,angle=0]{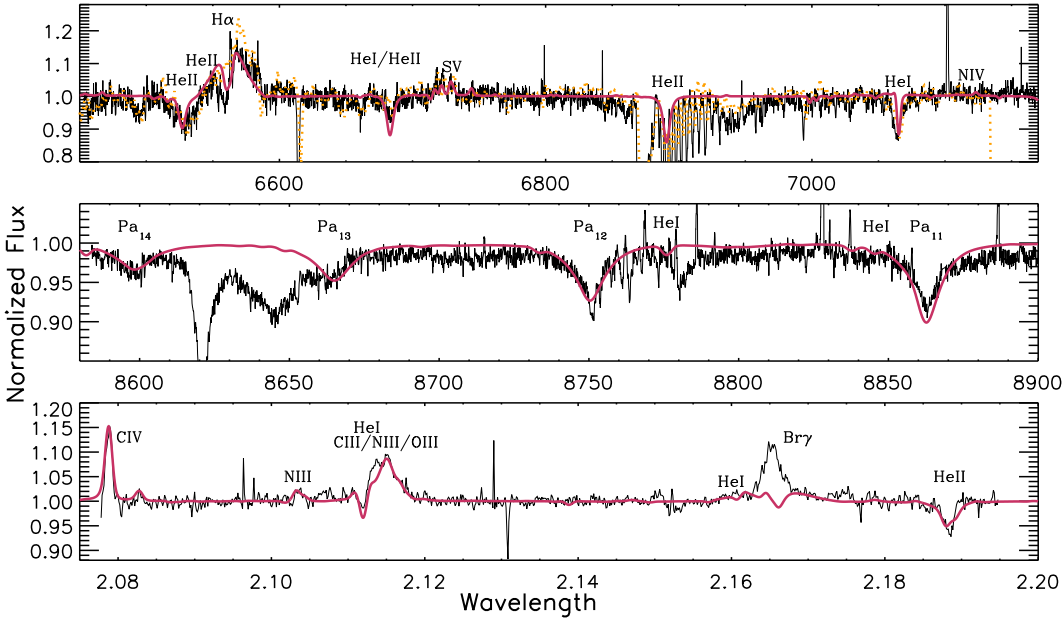}
\caption{Comparison of the synthetic spectrum of Wd1-30a derived from the optimised model-atmosphere solution described in 
Sect. 4 (red) to observational data (black). A further $R-$band spectrum is overplotted (orange) to 
demonstrate the night to night variability of the H$\alpha$ profile (spectra from 2004 June 12 \& 13).
The units of wavelength for the top and middle panels are Angstroms  and the bottom panel microns.}
\end{figure*}

\begin{figure*}
\includegraphics[width=14cm,angle=0]{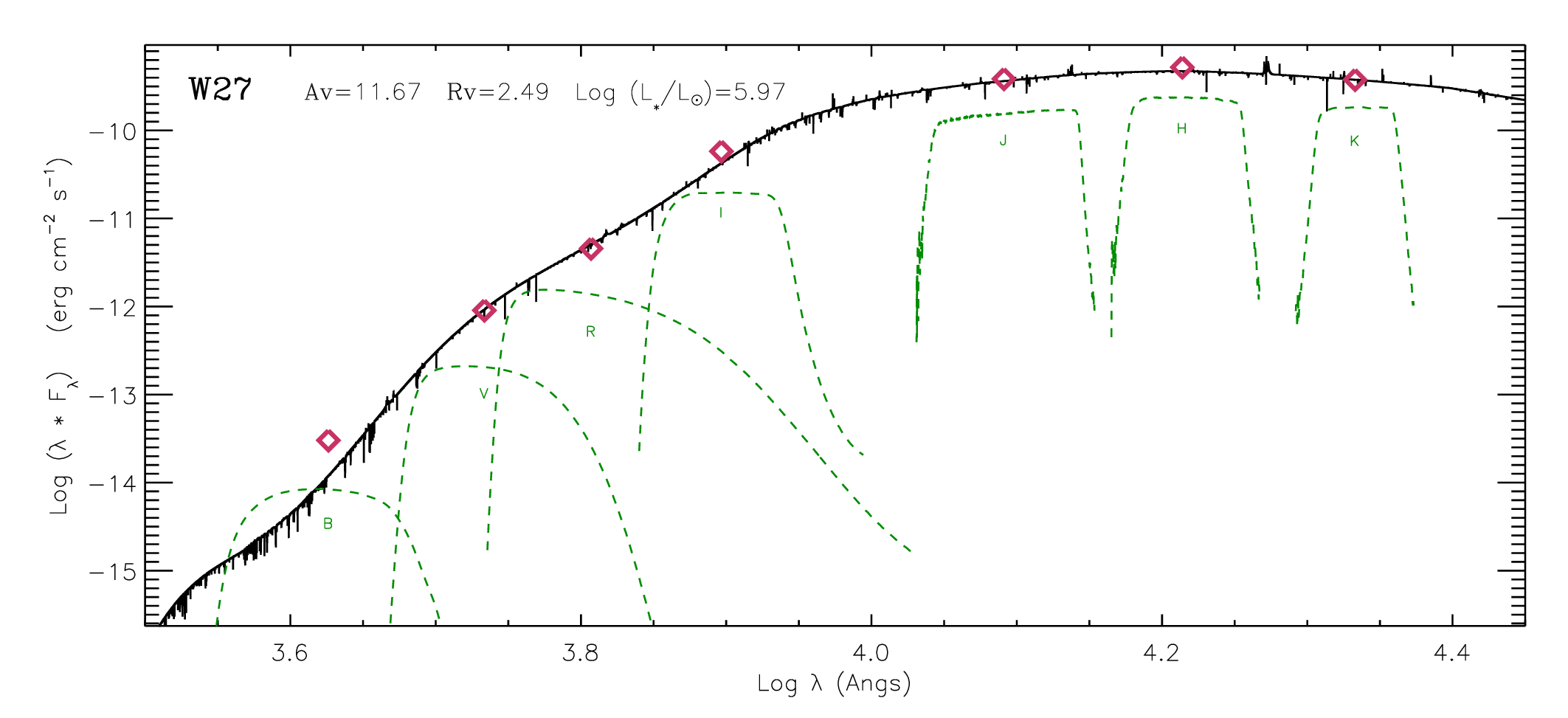}
\includegraphics[width=14cm,angle=0]{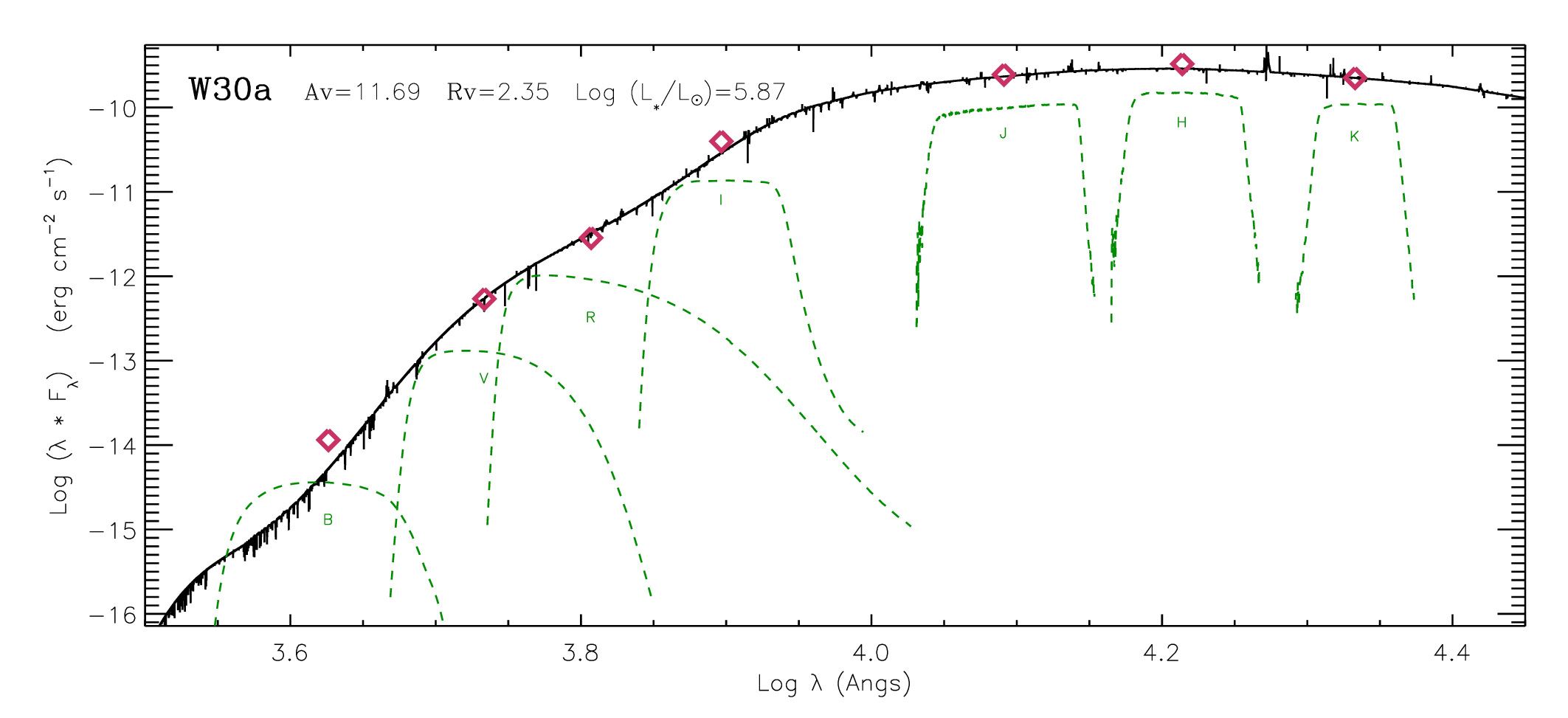}
\caption{Comparison of the synthetic spectral energy distributions  of Wd1-27 and -30a derived from model-atmosphere analysis 
to observations (Sect. 4). The band-passes of the filters are delineated by green dashed lines, photometry with red diamonds and the 
(reddened) synthetic spectra by the solid black line. Errors on photometric data are smaller than the symbol sizes. Reddening parameters and resultant bolometric luminosity are also indicated.}
\end{figure*}

\subsection{Spectral classification}
Turning first to the optical spectra of both Wd1-27 and -30a (Fig. 1) and the H$\alpha$ 
profiles in both stars appear somewhat  broader than those of the late-O/early-B 
supergiants that characterise Wd1 (Negueruela et al. \cite{iggy10}), being suggestive (at best)
of an earlier spectral type. Inflections in the profiles appear likely to be the result of P Cygni absorption
($\sim6558${\AA}) and He\,{\sc ii} photospheric absorption ($\sim6527.7${\AA} versus a rest wavelength of 6527{\AA}). 
The features  at $\sim6681.5${\AA} may represent blends of the He\,{\sc i} 6678{\AA} and He\,{\sc i} 6683{\AA}
 lines, while the strong  He\,{\sc i} 7065{\AA} photospheric line is present in both stars.

Moving to the I-band and the photometric Paschen series lines in both systems are seen to be anomalously weak  
in comparison to the majority of other late-O/early-B supergiants within Wd1 (e.g. Ritchie et al. \cite{ritchie09}); 
consistent with an earlier spectral classifiction (Fig. 1). However this phenomenon has also been observed in 
binaries, notably the  similarly X-ray bright system  Wd1-36  (Clark et al. \cite{clark15}, Ritchie et al. in prep.), although in 
this case the profiles of the stronger Paschen lines appear notably broader than seen in Wd1-27 and -30a, suggestive of 
contributions from two stellar components. With the possible exception of He\,{\sc i} 8777{\AA} and He\,{\sc i} 8847{\AA} 
in Wd1-27, none of the  other 
He\,{\sc i} photospheric  features (e.g. 8583{\AA} and  8733{\AA}) that characterise O9.5-B2.5 supergiants are present
(Ritchie et al. \cite{ritchie09}, Negueruela et al. \cite{iggy10}).  

However the $K-$band spectra are the most  diagnostically valuable. Specifically, the presence of strong C\,{\sc 
iv} 2.079$\mu$m emission and He\,{\sc ii} 2.189$\mu$m absorption unambiguously exclude classifications of O9 Ia or later 
(cf. Fig 2). Indeed, while weak He\,{\sc ii} 2.189$\mu$m absorption is present in the template spectra of O9-9.5 Ia  stars, it is  
absent for the remaining  supergiants within Wd1 for which K-band spectra are available. In combination with these features,  the 
presence of pronounced emission in the He\,{\sc i}+N\,{\sc iii} $\sim2.11\mu$m blend 
argues for a mid-O classification for Wd1-30a. Combined with weaker C\,{\sc iv} emission, the increased strength of   
He\,{\sc i} 2.112$\mu$m photospheric absorption in Wd1-27 suggests  a slightly later classification for that star. Finally, while the  
narrow Br$\gamma$ emission component seen in Wd1-27 is present in a subset of the  spectra of supergiants  over a wide range of 
spectral types (mid-O to early-B), the strong, pure emission line present in Wd1-30a is restricted to  high-luminosity early-mid O 
super-/hypergiants (Hanson et al. \cite{hanson05}, Clark et al. \cite{clark18a}, \cite{clark18b}). Indeed, employing the 
scheme utilised for the  Arches cluster (Clark et al. \cite{clark18a}) and $K-$band spectra from that work as classification templates we
assign O4-5 Ia$^+$ and  O7-8 Ia$^+$  for Wd1-30a and -27, respectively.

These classifications are consistent with the properties of  the optical spectra, but are substantially earlier than  the remaining 
late-O/early-B super-/hypergiant population of Wd1 which, as  can be seen from Fig. 2, are  well represented by the B0-1 Ia(b) 
K-band templates of Hanson et al. (\cite{hanson05}). Indeed, both Wd1-27 and -30a would appear to fit seamlessly into the stellar 
population of the Arches cluster (Fig. 2), which at $\sim2-3$Myr is significantly younger than Wd1 (Clark et al. 
\cite{clark05}, \cite{clark18a}). Finally, we see no evidence for a putative binary companion in the spectra of 
either star.

\subsection{Spectral variability}

Considering Wd1-27 first and the seven epochs of I-band spectra were searched for RV variability.
 The  RVs measurements were based on Levenberg-Marquardt nonlinear least-squares fits to Lorentzian profiles
for Pa11 and Pa13, with errors based on the internal statistical fit. No statistical robust RV variability was identified.
Both the Pa11 and Pa13 lines are well defined and of high S/N (cf. Fig. 3) and  the presence of a DIB at 8620$\AA$ 
provides  an excellent check for zero-point errors in the reduction and wavelength calibration; as a consequence  we consider 
this conclusion to be robust. Thus Wd1-27 would appear to be either a single star - potentially in tension with its
 X-ray properties - or a binary seen under an unfavourable inclination and/or with a low-mass companion.

Turning to Wd1-30a which,  as described in Negueruela et al. (\cite{iggy10}),  appears spectroscopically  
variable. Specifically  the emission line profile of H$\alpha$ appears changeable on short ($\sim$day) 
timescales (Fig.4), while in the I-band data there is also the  suggestion of He\,{\sc ii} 8236{\AA} in some 
spectra and changes in the strength of the C\,{\sc  iii} 8500{\AA} line. Examining the I-band data and despite 
the weak Paschen series lines we can rule out the null hypothesis (that there are no RV variations) at
$>99$\% confidence ($>4\sigma$). Even if we arbitrarily  double the errors, we still reject the null hypothesis at
$>99$\% confidence. We find a systemic velocity of $-39\pm2$kms$^{-1}$; fully consistent with the cluster 
mean (cf. Clark et al. \cite{clark14}, in prep.); supportive of  the conclusion that our RV measurements are accurate.
We find a low semi-amplitude for the variability of $\Delta$RV$\sim12{\pm}3$kms$^{-1}$. The standard deviation in RVs is 
$\sim8$km s$^{-1}$; this is  approximately twice the mean internal error,  which implies that determining a unique period(s) from 
these data will be difficult. 

Such behaviour could derive from either stellar pulsations or orbital motion. Unfortunately the low semi-amplitude 
of the variability precludes searches for multiple pulsational modes in these data such as are observed in other cluster 
members (cf. Wd1-71; Clark et al. in prep.). However observations of 
the wider cluster population (Ritchie et al.   \cite{ritchie09}; see also Sim\'{o}n-D\'{i}az et al. \cite{simon}) suggests that 
pulsations become apparent around spectral  type  $\sim$B0.5-1 Ia   as the stars start to evolve redwards and consequently the 
earlier spectral type of Wd1-30a might  be taken as circumstantial evidence  that orbital motion is a more likely source of the 
RV variability; a conclusion seemingly well-supported by its X-ray properties. 

As a consequence a period search was carried out
using the reference implementation of the
fast~$\chi^2$ algorithm (Palmer \cite{palmer}) with a
single harmonic component. A number of periodicities of $\leq10$ days were returned, 
with the strongest peak in the resultant 
periodogram found at $\sim4.5$days. This was found not to be statistically 
significant in comparison to other periods returned, while additional peaks in the periodogram 
above $\sim10$days were found to be aliases of shorter periods.
 With limited sampling and a low semi-amplitude, false periods arising
from noise and aliasing are problematic, and $k-1$ cross-validation
was therefore used to assess the periodicities found. In
each resampled subset false peaks in the periodogram are expected to  
vary while a peak due to the true orbital period will remain constant,
providing some confirmation that the true period has been
identified. Unfortunately, none of the periods returned in our analysis
appear robust to such verification.

 However, we do see a statistically-significant RV shift from $-28.3\pm2.6$kms$^{-1}$
 on MJD 54665.03556 to $-42.8\pm2.3$kms$^{-1}$ on MJD 54671.13430 (6.099 days later), which supports the inference  of a short 
orbital period for Wd1-30a. Such a conclusion would be fully consistent with 
scenarios whereby  Wd1-30a is either (i) a pre-interaction system with a rather extreme mass-ratio and/or is observed under 
unfavourable inclination or (ii) a post-interaction system that is now 
dominated by the mass-gainer,  which shows relatively low RV shifts as a result of a greatly reduced mass for the original
 primary star. 
 However the former hypothesis  would potentially struggle to explain both the X-ray properties of Wd1-30a, which appear 
indicative of a massive CWB (Sect. 1), and the anomalously early spectral type inferred for it (Sect. 3.1.1).

\subsection{Cluster membership}

 Given the properties of the wider cluster population, the spectral types reported above for Wd1-27 and 
-30a appear anomalously  early. An obvious question is therefore whether they are {\em bona fide} cluster 
members or instead  interlopers located along  the line of sight to Wd1. Their 
apparent magnitudes and  reddenings are certainly consistent with cluster membership (e.g. Negueruela et 
al.  \cite{iggy10}, Sect. 4), while their X-ray properties, while extreme,  are also within the envelope defined for 
Wd1 by other massive OB supergiant  and WR binaries (Clark et al. \cite{clark08}). As highlighted in the 
previous subsection, the systemic radial velocities of both systems are directly comparable to the cluster 
mean value.

Critically however, we  may employ analysis of the second Gaia data release (DR2; Gaia Consortium et 
al. \cite{gaia16}, \cite{gaia}) to help address this question for these and, indeed,  other putative cluster members.
For reasons of continuity   we present the detailed analysis of these data in Appendix A, simply 
summarising the essential points here.   Upon consideration of the parallaxes reported, it is immediately 
obvious  that these may not, in isolation, be utilised to  determine the distance 
to individual stars within Wd1. As a consequence, after an  initial photometric colour 
cut to exclude foreground objects, we utilised a  combination of parallax 
and proper motion data to construct an astrometrically  defined cluster 
population. We find the properties of both Wd1-27 and -30a to be  
consistent with cluster membership on this basis.

\section{Quantitative modeling}

\begin{table*}
\begin{center}
\caption{Model parameters for Wd1-27.}
\label{tab:model}
\begin{tabular}{ccccccccccccc}
\hline\hline
  log($L_*$) & $R_*$     & $T_{\rm eff}$  & ${\dot M}$             & $v_{\rm \infty}$ & $\beta$ &  $f_{\rm cl}$  & log$g$   & $M_*$ & He/H & N/N$_{\odot}$ & C/C$_{\odot}$ & O/O$_{\odot}$ \\

  ($L_{\odot}$)       &  ($R_{\odot}$)     &  (kK)    &  ($10^{-6} M_{\odot}$yr$^{-1}$) & (kms$^{-1}$)   &         &           &  &  ($M_{\odot}$)   & &   & \\
\hline
     5.97$_{-0.10}^{+0.15}$     &  28.5$^{+3.0}_{-3.0}$      & 33.5$^{+1.5}_{-1.5}$   &   3.10$^{+0.47}_{-0.47}$             &  2200$^{+300}_{-700}$ &  1.0$^{+0.50}_{-0.15}$  & 0.075 &   3.38$^{+0.15}_{-0.10}$  & 71.3$^{+38.0}_{-27.0}$  & 0.175$^{+0.075}_{-0.05}$    & 10.9$^{+5.0}_{-5.0}$       
&   0.33$^{+0.16}_{-0.16}$          &   0.52$^{+0.26}_{-0.26}$        \\
\hline
\end{tabular}
\end{center}
{We adopt a distance of $\sim5$~kpc to Wd1 (Negueruela et al. \cite{iggy10}); consistent with the Gaia parallaxes for cluster members (Appendix A). Errors on the stellar luminosity and  radius assume the cluster distance is well determined and derive from uncertainty in the correction for interstellar reddening (Sect. 4.1.2) and stellar temperature. The error on the spectroscopic mass is derived from  the propagation of (assumed) Gaussian uncertainties.
We note that $R_*$ corresponds to $R(\tau_{\rm Ross}=2/3)$. The H/He ratio is given by number and other abundances are relative to solar
values
from Anders \& Grevesse (\cite{anders});
if we use the values from Asplund et al. (\cite{asplund}) as a reference, the derived ratios need to be
scaled by 1.38, 1.537 and 1.86 for C, N and O respectively. Note that $f_{\rm cl}$ is derived for the inner, line-forming regions of the stellar wind and not for the outer regions responsible for the mm-continuum; see Sect. 4.2 for further details. }
\end{table*}

\begin{table*}
\begin{center}
\caption{Model parameters for Wd1-30a.}
\label{tab:model}
\begin{tabular}{ccccccccccccc}
\hline\hline
  log($L_*$) & $R_*$     & $T_{\rm eff}$  & ${\dot M}$             & $v_{\rm \infty}$ & $\beta$ &  $f_{\rm cl}$  & log$g$   & $M_*$ & He/H & N/N$_{\odot}$ & C/C$_{\odot}$ & O/O$_{\odot}$ \\

  ($L_{\odot}$)       &  ($R_{\odot}$)     &  (kK)    &  ($10^{-6} M_{\odot}$yr$^{-1}$) & (kms$^{-1}$)   &         &           &  &  ($M_{\odot}$)   & &   & \\
\hline
     5.87$_{-0.10}^{+0.15}$     &  20.6$^{+3.0}_{-3.0}$      & 37.25$^{+1.0}_{-2.5}$   &   1.33$^{+0.20}_{-0.20}$             &  1200$^{+200}_{-400}$ &  1.15$^{+0.65}_{-0.05}$  & 0.035 &   3.65$^{+0.15}_{-0.10}$  & 69.4$^{+37.0}_{-26.0}$ & 0.2$^{+0.2}_{-0.05}$    & 10.9$^{+5.0}_{-5.0}$       
&   0.93$^{+0.46}_{-0.46}$          &   0.84$^{+0.42}_{-0.42}$        \\
\hline
\end{tabular}
\end{center}
\end{table*}

\subsection{Methodology}

In order to determine the underlying physical parameters of Wd1-27 and -30a we employed the non-LTE model-atmosphere code
CMFGEN (Hillier et al. \cite{hillier98}, \cite{hillier99}) in a two-stage process. Initially we just employed the spectroscopic datasets for both stars and utilised a $\chi^2$ minimisation technique to find the best-fits to them  from a grid of $\sim3500$ unique models sampling the parameter space suggested by previous analysis of stars of comparable spectral type and luminosity class (cf. Najarro et al. \cite{najarro04}).
Specifically, the grid well samples primary physical parameters such as $T_{\rm eff}$, log$g$, He abundance and wind density and allows reliable estimates of their uncertainties from our $\chi^2$ fitting. By computational necessity the rest of the parameters are more sparsely sampled, which unfortunately does not allow for a simultaneous determination of the uncertainties for all physical properties. Instead, errors for metal-abundances (C, N, O and Si), $v_{\rm \infty}$ and 
 $\beta$, were estimated by means of smaller grids where we let these parameters vary after fixing the primary ones.
 Each spectral diagnostic line (Sect. 4.1.1) was given a weight for the fitting which was set by the S/N of the spectral region 
 in question. Special weighting was also applied to investigate the role  of diagnostics in specific bands ($R-$, $I-$ or $K-$ 
 band)  in constraining the stellar properties, of particular relevance to the evaluation of Wd1-30a (Sect. 4.3); however we 
emphasise that final fitting was accomplished via the former methodology.

 Spectroscopic  modeling yields all stellar properties (cf. Tables 2 and 3)  except the absolute values for bolometric  
luminosity, mass-loss rate and stellar radius, which can be obtained by the application of an appropriate scaling factor determined 
via fitting the model spectral energy distribution (SED) to the  observed photometry after accounting for distance (assumed to be 5kpc) and interstellar extinction (Sect. 4.1.2). To accomplish this  a Marquardt-Levenberg technique was applied to obtain the best fitting model. 
 Once the final model was constructed, the 3mm continuum flux was obtained from the scaled SED in order to
confront it with the ALMA observations.

\subsubsection{Spectral diagnostics}
 We employed a large number of line diagnostics in order to constrain the  bulk properties of both stars.
\begin{itemize}
\item{{\bf Temperature:} 
The He\,{\sc i} 6678$\AA$/He\,{\sc ii} 6683$\AA$ ratio is a prime diagnostic in the $R-$band, but 
in the parameter domain in question the He\,{\sc i} 6678$\AA$  component is unfortunately extremely sensitive 
to the effect of the Fe\,{\sc iv} extreme ultraviolet (EUV) lines which overlap with the He\,{\sc i} singlet
 1s$^2$ $^1$S - 1s2p$^1$1P$^{\rm o}$ resonance line at 584.334$\AA$ (cf.  Najarro et al \cite{paco06}). Nevertheless, and subject to S/N,
 the apparent lack of He\,{\sc i} 6678$\AA$ sets a  lower limit on $T_{\rm eff}$ for both stars. Similarly the 
S\,{\sc v} lines at 6717, 6722 and 6729{\AA} also provide a lower bound, while the absence  of the N\,{\sc iv} 7103-7122 lines
 provides an upper limit to  $T_{\rm eff}$.

In the $K-$band the He\,{\sc i} 2.112$\mu$m triplet absorption component is a `classical' diagnostic for spectral type/$T_{\rm eff}$ 
for O stars. However for O If$^+$ stars, where  Br$\gamma$ is typically seen in emission, the
He\,{\sc i} 2.112$\mu$m transition demonstrates an additional dependence on mass-loss and surface gravity. As such, inspection of the spectra (Fig. 2) suggests that while it  constitutes a good diagnostic for $T_{\rm eff}$ in Wd1-27, it only provides a lower limit for Wd1-30a.  The He\,{\sc i} 2.113$\mu$m singlet  component displays a similar behavior to He\,{\sc i} 6678$\AA$ (via EUV coupling), although, for  $T_{\rm eff}<34$kK, it plays a significant role in the 2.113-2.116$\mu$m emission blend. In this domain He\,{\sc ii} transitions show a strong temperature dependence, with the  He\,{\sc ii} 2.189$\mu$m line serving  as a prime diagnostic. Finally the C\,{\sc iv} lines also show a temperature dependence, although their sensitivity to other physical properties (mass-loss, C abundance, velocity field and surface gravity) limit their utility.}
 
 \item{{\bf Surface gravity:} Despite the uncertainty induced by the 
 placement of the continuum and the presence of multiple DIBs, the 
 detailed shape of the photospheric Paschen absorption lines in the $I-$ 
band provide a valuable constraint on log$g$. }

\item{{\bf Elemental abundances:} Turning first to  helium and  the strength of the He\,{\sc ii} 6527, 6683, 6890$\AA$ and 
2.189$\mu$m lines may all be 
utilised, although poor S/N and tellurics/DIBs detract from  6527 and 6890$\AA$,  while the powerful wind of Wd1-30a affects the 
2.189$\mu$m transition. For He\,{\sc i} one can employ the 7065$\AA$ transition (although located within a telluric region), 
the lines at 8733, 8776 and 8845$\AA$ in the $I-$band and 2.112$\mu$m and the He\,{\sc i} 7-4 transitions around Br$\gamma$ in the $K-$band
 (though the latter  are also strongly dependent on the turbulent velocity and clumping structure).

For nitrogen our primary diagnostic is the N\,{\sc iii} 2.1035$\mu$m line, with the N\,{\sc iii} 2.155$\mu$m transition that is blended with C\,{\sc iii} and O\,{\sc iii} also useful. Given the high temperature regime considered, the N\,{\sc iv} lines between 
7103-7122$\AA$ may also be  employed as secondary diagnostics.

Despite their dependence on multiple stellar prarameters we are forced to employ the $K-$band C\,{\sc iv} transitions - with the abundance fixed only after all other physical properties are set - although a weak C\,{\sc iii} line at 2.11$\mu$m may be used as a secondary diagnostic.

 The O\,{\sc iii} 8-7 transitions dominate the red part of the 2.115 broad emission feature and  can be  
 used to constrain the oxygen abundance to within 0.2 dex, while the 6-5 O\,{\sc iii} lines in the $I-$band also provide an upper limit. 

Finally, despite being located within a noisy, strong telluric region of the spectrum, we are forced to employ the Si\,{\sc iv} line at 8957{\AA} to determine the silicon abundance, although due to its additional  dependence on both $T_{\rm eff}$  and
 log$g$, it may only be determined once the rest of the parameters are set.}

\item{{\bf Wind properties:} The shape and strength of the profiles of 
both H$\alpha$ and  Br$\gamma$ are extremely sensitive to mass-loss rates (\.{M}), wind clumping and the 
velocity law. He\,{\sc ii} 2.189$\mu$m can also be used as a secondary diagnostic for mass-loss and clumping, 
especially for the O If$^+$ models, where Br$\gamma$ is in emission.}
\end{itemize}

\subsubsection{Interstellar extinction}

For stars suffering significant interstellar reddening, application of the
correct extinction law is essential if reliable physical
parameters are to be returned. The Arches and Quintuplet clusters
illustrate this issue, with differences in bolometric luminosities
for cluster members of up to $\sim0.6$dex being returned depending on
the extinction law employed (Clark et al. \cite{clark18a},
\cite{clark18b}). As a consequence significant effort was employed in
testing a number of differing models, an identical (but expanded)
approach to that adopted for analysis of Wd1-5 (Clark et al. \cite{clark14}).
The following reddening laws were compared  \footnote{We did not make use of
the tailored recent formulation of Hosek et al. (\cite{hosek}) as it does not
extend to optical wavelengths.}:
 \begin{itemize}
 \item{Our current tailored prescription consisting of the prescription
 provided by Cardelli (\cite{cardelli})
 below 1$\mu$m, a modified Rieke \& Lebofski (\cite{rieke}) law for the
1.0-2.5$\mu$m range and the Moneti et al. (\cite{moneti}) formulation
 for longer wavelengths.}
 \item{A family of NIR-$\alpha$ laws of the kind $A_{\lambda} = A_{Ks}
 ( \lambda_{Ks} / \lambda )^{\alpha}$ for a number of different power-law
 indices ($\alpha\sim1.53-2.32$)}
 \item{The extinction law by   Ma\'{i}z Apell\'aniz et al. (\cite{MA14}). }
 \item{The Fitzpatrick (\cite{FM99}) prescription.}
 \item{The optical-NIR law specifically constructed for Wd1 by Damineli et al.
 (\cite{damineli}).}
 \end{itemize}

 Our extinction law provides the best fit (minimum $\chi^2$) to the
  optical (SUSI) and NIR (SOFI) photometry. The same results were obtained
  when replacing our optical SUSI data by those of Lim et al.
  (\cite{LIM13}), which were utilized by Damineli et al. (\cite{damineli}).
  Contrasting the different laws shows that simple power-law
  formulations with canonical exponents
  yielded systematically fainter bolometric luminosities. Similar
  behaviour was found for members of the Arches and Quintuplet, where
  unphysically low luminosities were returned  (Clark et al.
  \cite{clark18a}, \cite{clark18b}).  Indeed, while single power laws
  may be appropriate for the near-IR (1$\mu$m $< \lambda < 2.5\mu$m),
  they clearly fail when extended to optical wavelengths (cf. Fig. 4
  of Damineli et al. \cite{damineli}). Our extinction law reveals
   $R_v\sim2.2-2.5$  and $A_v\sim11.7$ for both stars and yields
   results that are broadly comparable to  those returned by application
   of the Fitzpatrick (\cite{FM99}) and  Ma\'{i}z Apell\'aniz et al.
   (\cite{MA14}) prescriptions, with the bolometric luminosities resulting from the latter
two formulations less than $0.1$dex in excess of those reported in Sect. 4.2; this close 
equivalence provides confidence in our  approach. Conversely,
   while we obtain similar $A_v$ values to Damineli et al.
   (\cite{damineli}) for both objects, their $A_{Ks}\sim0.73$ are slightly
   lower than ours ($A_{Ks}\sim1.2$) implying lower luminosities ($\sim
   0.2$dex)   if their extinction law is used\footnote{Note that if this 
law were adopted then previous luminosity (and consequently mass) determinations (e.g. Negueruela et al. \cite{iggy10},
Clark et al. \cite{clark14}) used for comparison to Wd1-27 and -30a would also have to be 
systematically revised downwards; thus the overal  conclusions of this paper would still be valid
(e.g. Sects. 4.2, 5 \& 6).}.

.

\subsection{Wd1-27 results}

The results of modeling Wd1-27 are presented in Table 2, with comparison of (non-simultaneous) observational data to the synthetic spectrum and SED in Figs. 3 and 5. An excellent match is found to both spectroscopic and photometric data, with the sole exception of the emission component present in Br$\gamma$; nevertheless with H$\alpha$ in emission and Br$\gamma$ substantially infilled the wind properties are well defined. The stellar temperature is consistent with our spectral classification, while the elemental abundances are consonant with this picture, being indicative of  moderate CNO processing. Both bolometric luminosity and  spectroscopic mass (ultimately derived from the high S/N Paschen series lines) are surprisingly high for cluster members, a finding we return to in Sect. 5. 
We predict a 3mm flux of $\sim0.15$mJy; slightly larger than the 0.13mJy $3\sigma$ observational upper limit,
 which implies a wind-clumping factor, $f_{\rm cl}\geq0.4$ at radii where  the mm-continuum arises;  larger 
than inferred for the line forming regions and hence suggestive of  a radial dependence to this property\footnote{It is expected that both optical and near-IR line formation regions will be physically distinct and much closer to the photosphere than the zone in which the 
3mm continuum arises. Hence  the 3mm flux may be matched by
 simply scaling the clumping in the outer wind without altering the $R-$,$I-$ and $K-$band spectra (cf. Najarro et al. \cite{paco11}).}.

We utilised the Bonnsai tool (Schneider et al. \cite{schneider14b}) to compare these parameters to the evolutionary models of Brott et al. (\cite{brott}) in order to  infer an age and initial mass for Wd1-27. 
Utilising the values of $L_{\rm bol}$, $T_{\rm eff}$, log$g$ and surface helium abundance from Table 2 Bonnsai returns an age of 
2.6Myr and initial and current masses of  $65.8M_{\odot}$ and $54.6M_{\odot}$ respectively. 
As expected for the  current mass predicted by  Bonnsai, the value of log$g\sim3.35$ is slightly smaller than found via 
modelling, although within the uncertainty on this parameter(Table 2). Intriguingly, Bonnsai did not return the surface helium 
abundance found by modelling.   Foreshadowing Sect. 5 we suspect that both Wd1-27 and -30a have experienced significant binary 
interaction, which  has led to the  anomalously high He-abundances suggested by our analysis and may in turn explain the 
resultant discrepancy between our modeling and the predictions of Bonnsai, which assumes a single-star evolutionary channel.

\subsection{Wd1-30a results}

Modeling results for Wd1-30a are presented in Table 3 and Figs. 4 and 5. Unlike Wd1-27, while we reproduce the SED we fail to simultanously replicate the $R-$, $I-$ and $K-$band spectroscopy. Comparison of our best-fit synthetic spectrum to observational data shows that while most spectral diagnostics are well fit, we fail to duplicate the Br$\gamma$ emission profile - despite success with H$\alpha$ and the Paschen series lines - and the bluewards emission in the 2.11$\mu$m blend (attributed to
He\,{\sc i} 2.112$\mu$m). It is possible to fully replicate the $K-$band spectrum, including Br$\gamma$, but at the cost of grossly over-estimating the strength of
 H$\alpha$ emission and depth of the photospheric Paschen series lines (Fig. B.1). How might we explain this discrepancy? We highlight that the differing spectra (and photometry) are non-contemporaneous and furthermore that the star is clearly spectroscopically variable (Sect.  3.2 and Fig. 4); hence we consider it most likely that the $K-$band spectrum was obtained at an epoch (or orbital phase) in which mass-loss was temporarily enhanced. Clearly further  simultaneous spectroscopic observations will be necessary to test this hypothesis\footnote{One might assume that we have erroneously observed different stars in the R- and K-bands. We consider this unlikely given the IR spectroscopy of other cluster members matches that expected from optical observations; hence one would have to invoke this solely for Wd1-30a. Moreover, if the K-band spectrum did not correspond to Wd1-30a
it would imply the presence of an additional star of anomalously early spectral type within the core of Wd1, which would have to be an exceptionally  strong H$\alpha$ emitter. No such star has been identified in any of our long- and multi-slit spectroscopic programs (e.g. Clark et al. \cite{clark05}, Negueruela et al. \cite{iggy10}), nor in our NTT/EMMI and VLT/FORS slitless spectroscopy of the whole cluster  (e.g. Negueruela \& Clark \cite{iggy05}) or the narrow-band H$\alpha$ imaging employed by  Wright et al. (\cite{wright14}).
And even if such an hypothetical  star were to be discovered, its presence would still be consonant with the central conclusion of this paper, namely that it and Wd1-27 - for which spectroscopic modeling is consistent and unambiguous - are of earlier spectral type and more massive and luminous  than expected given the remaining stellar population of Wd1.}.

Nevertheless proceeding under this scenario and, mirroring Wd1-27, Wd1-30a is an hot, highly luminous and massive star with a powerful (albeit slower) wind; again consonant with expectations from our spectroscopic classification. The predicted 3mm-continuum flux of 0.19mJy  is consistent with observations ($0.17\pm0.06$mJy) assuming only a minor evolution in the clumping factor between the line-
($f_{\rm cl}\sim0.035$) and continuum-forming ($f_{\rm cl}\sim0.085$) regions of the wind. Finally we note that nitrogen appears 
anomalously enhanced  given that both carbon and oxygen seem barely depleted. 

Following Sect. 4.2 Bonnsai predicts an age of 2.3Myr and  initial and final masses of  $56.6M_{\odot}$ and $51.0M_{\odot}$ 
respectively; as with Wd1-27 the latter being lower than our spectroscopic estimate. Likewise, 
 the surface helium abundance was not returned and  the value of log$g\sim3.56$ was also lower than anticipated, although just 
within the uncertainty on this parameter (Table 3).

\section{Discussion}

Both qualitative classification and quantitative analysis suggest that Wd1-27 and -30a are very massive mid-O hypergiants.
Indeed, the bulk
 properties of both stars ($L_{\rm bol}$, $M_*$, $T_{\rm eff}$, \.{M} and $V_{\infty}$) are  directly comparable to other known Galactic  examples\footnote{The secondary in Arches F2 (Lohr et al. \cite{lohr}),
the primary in Cyg OB2 B17 (Stroud et  al. \cite{stroud}) and the primary of the X-ray binary HD153919 (Clark et al. \cite{clark02}).}, corroborating our analyses and conclusions. 
 Under the assumption that both objects  evolved via a single star channel, the Bayesian evolutionary tool Bonnsai (Schneider et al. \cite{schneider14b}) predicts that both stars are 
young  (2.3-2.6Myr) and of  high initial and current mass, although we caution that it is unable to replicate all physical parameters derived from quantitative modeling.

In contrast, the remaining OB supergiant(hypergiant) population of Wd1 demonstrate spectral types ranging from O9-B4 Ia (B0-B9 Ia$^+$); 
consistent with a cluster age of $\sim5$Myr inferred from consideration of the complete stellar census of both hot and cool stars (Clark et al.  \cite{clark05}, \cite{clark15}, Crowther et al. \cite{crowther06}, Negueruela et al. \cite{iggy10}). Quantitative model-atmosphere analysis has yet to be performed for the majority of cluster members, but construction of a semi-empirical HR-diagram and application of appropriate bolometric corrections according to spectral type suggests log$(L_{\rm bol}/L_{\odot})\sim5.7-5.8$ for the OB supergiants and mid- to late-B hypergiants (Clark et al. \cite{clark05}, Negueruela et al. \cite{iggy10}). Similar luminosities are also suggested for the cooler yellow hypergiants within the cluster (Clark et al. \cite{clark05}). Indeed only the luminous blue variable Wd1-243 is of apparently higher luminosity, with quantitative analysis by   Ritchie et al. (\cite{ritchie09b}) suggesting log$(L_{\rm bol}/L_{\odot})\sim5.95$; however such stars are known  bolometric luminosity variables (e.g. Clark et al. \cite{clark09} and refs. therein).

 On the basis of their spectral classifications, temperatures for the  population of early-type stars within Wd1 range from 32kK 
 for the O9 Ib-III stars, through to 28kK for the B0 Ia cohort and $\sim13.5$kK for the mid-B hypergiants (Negueruela et al. 
 \cite{iggy10}). As expected from their spectral types, the temperatures determined for  Wd1-27 and -30a from modelling are in 
 excess of these estimates. Evolutionary masses of $\sim35M_{\odot}$ are inferred for the wider OB supergiant population 
 (Negueruela et al. \cite{iggy10}); consistent with the analysis of Crowther et al. (\cite{BSG}) who find evolutionary masses for 
 galactic B0-3 supergiants to range from $\sim25-40M_{\odot}$. Furthermore we are fortunate to be able to determine a {\em current} 
 dynamical mass estimate of $35.4^{+5.0}_{-4.6}M_{\odot}$  for the OB supergiant within the eclipsing binary Wd1-13 (Ritchie et 
 al. \cite{ritchie10}). Both results suggest an upper limit of $\sim40M_{\odot}$ for the OB supergiants within Wd1; even allowing 
 for the large formal uncertainties on the mass estimates for Wd1-27 and -30a they are both still in excess of this value. 
 Moreover, as described above, the consonance between our spectroscopic masses and the dynamical mass  estimates for other 
 galactic mid-O hypergiants gives confidence in the conclusion that Wd1-27 and -30a are indeed signficantly more massive than the 
 cluster supergiants.

To summarise - comparing the above properties to those of Wd1-27 and -30a (Tables 2 and 3) reveals the latter to be  hotter, more luminous and massive  than the other  members of Wd1 and, as a consequence, apparently younger\footnote{If we were to apply the Damineli et al. (\cite{damineli}) extinction law (Sect. 4), Bonnsai returns initial masses of $47.2M_{\odot}$ and $44.4M_{\odot}$ and 
ages of $\sim2.8$Myr and $\sim2.6$Myr for Wd1-27 and -30a, respectively (assuming single star evolution). Therefore the conclusion that the stars are more massive and younger than the remaining cluster members appears robust.}.
As such they fulfill the classical definition of blue stragglers. The origin of low-mass blue stragglers has been the 
subject of much discussion, with mass transfer or merger in binaries (e.g. McCrea \cite{mccrea}) and stellar collisions 
(e.g. Hills \& Day \cite{hills}) seen as the prime formation channels, although the relative weighting of
 each is still uncertain (cf. review by Knigge et al. \cite{knigge}).  

With the additional information afforded by the RV observations and quantitative modeling, it is appropriate to revisit the nature of both Wd1-27 and -30a. As highlighted in Sect. 1, Wd1-27 has an X-ray flux and spectrum consistent with the short-period binaries Wd1-26 and -53. However the X-ray luminosity is within a factor of two of expectations given its extreme bolometric luminosity, with recent work by Nebot G\'{o}mez-Mor\'{a}n  \& Oskinova (\cite{nebot}) suggesting considerable scatter in the $L_{\rm X}/L_{\rm bol}$ ratio. Given this, and in the absence of RV shifts, we cannot at present determine whether  Wd1-27 is a binary or single star and hence whether it formed via mass transfer or merger, respectively.

 In contrast Wd1-30a has an X-ray flux a factor of five greater than expected and, critically, a spectrum considerably harder than predicted for a single star (Sect. 1). In conjunction with  the presence of RV variability (with an apparent periodicity of $\leq10$ days) we consider it likely that it was initially the  secondary in a massive binary system that has subsequently experienced extensive  mass transfer, such that it now dominates emission from the system. The chemical abundances derived for Wd1-30a support such a scenario (Table 3), with the enhancement of nitrogen in the absence of carbon and oxygen depletion inconsistent with expectations for  the exposure 
of nuclear burning products at the stellar surface due to rotational mixing. Instead, one might suppose they result from the transfer of significant quantities of mass due to binary interaction (e.g. Hunter et al. \cite{hunter08}, \cite{hunter09}, Langer et al. \cite{langer08}).

\subsection{Stellar evolution in Wd1}

One of the most striking findings derived from this work is that Wd1 possesses a uniquely rich  population of both hot 
and cool super-/hypergiants, extending from early/mid-O spectral types (O4-5 Ia$^+$; Wd1-30a) through to the most luminous red 
supergiants known (M2-5 Ia; Wd1-26). To the best of our knowledge no other cluster replicates this range and no models for single 
star evolution in a co-eval cluster can reproduce this distribution (e.g. Brott et al. \cite{brott}, Ekstr\"{o}m et al. 
\cite{ekstrom}). 

Binary interaction potentially  offers an explanation for this observation. Previously two evolutionary channels had been inferred 
to 
operate within Wd1 (Clark et al. \cite{clark11}, \cite{clark14}). The first, for single stars and long period binaries is 
responsible for the formation of the B5-9 hypergiants (Wd1-7, -33 and -42) and their progression through yellow hypergiant 
(e.g. Wd1-4, 12a and  16a) and red supergiant (Wd1-20, 26 and 237) phases prior to evolving bluewards across the HR diagram to 
become H-depleted WRs. The second envisages stripping of the outer layers of the primary in a close binary system undergoing 
case-A/early case-B mass transfer. Examples of this evolutionary channel would be Wd1-13 (Ritchie et al. \cite{ritchie10}) 
and the putative binary which contained Wd1-5 (Clark et al. \cite{clark14}) and it yields a population of 
chemically peculiar early-B hypergiant/WNVLh stars that are overluminous for their current mass, but underluminous in comparison to other cluster BHGs\footnote{Wd1-5 has log($L_{\rm bol}/L_{\odot})\sim 5.38$ and Wd1-13  log($L_{\rm bol}/L_{\odot})\sim5.2$ (Ritchie et al. \cite{ritchie10}, Clark et al. \cite{clark14}) compared to  log($L_{\rm bol}/L_{\odot})\sim5.97$ for Wd1-27, log($L_{\rm bol}/L_{\odot})\sim5.87$ for Wd1-30a and log($L_{\rm bol}/L_{\odot})\sim5.8$ for the apparently single B5 Ia$^+$ hypergiants Wd1-7 and -33 (Neguerueula et al. \cite{iggy10}).}. These will subsequently avoid a red loop across the HR-diagram, 
remaining at high temperatures and evolving directly to the WR phase.  

Finally our results support the (third) channel suggested by de Mink (\cite{demink}) and Schneider et al. (\cite{schneider14}) in 
which quasi-conservative mass-transfer in a close binary leads to the  rejuvenation of the secondary via mass transfer or merger. The nature of the resultant 
binary product obviously depends on the initial mass of the secondary and the quantity of mass transferred, but at this instance  
leads to the production of early-mid O hypergiants within Wd1 (implying  that  three distinct evolutionary 
pathways can lead  to the formation of blue hypergiants)\footnote{While there is no evidence for binary evolution  in the Arches 
(Clark et al. \cite{clark18a}), the presence of both O7-8  Ia$^+$ and B0-3 Ia$^+$ hypergiants within the Quintuplet suggest that both single 
and binary channels are in operation in that cluster (Clark et al. \cite{clark18b}; Sect. 5.2).}.

Unfortunately the efficiency at which matter may be accreted by the secondary in a close binary  is highly uncertain 
(although this will not affect the evolution of the primary). Simulations  by  Petrovic et al. (\cite{petrovic}) suggest that 
mass-transfer is highly non-conservative once the secondary has been (quickly) spun-up to 
critical rotation; the evolutionary calculations for Wd1-13 by Ritchie et al. (\cite{ritchie10}) were undertaken under this 
assumption. Conversely Popham \& Narayan (\cite{popham}) suggest that sufficient angular momentum may be lost via an accretion 
disc around the secondary that it may continue to accrete even if at critical rotation, leading to quasi-conservative mass 
transfer. This is a significant 
difference over the approach of Petrovic et al. (\cite{petrovic}) since it favours pronounced rejuvenation of the secondary. The 
evolutionary code employed by de Mink et al. (\cite{demink}) and the cluster population synthesis studies of Schneider et al. 
(\cite{schneider14}, \cite{schneider15}) employ this formulation, and it was utilised to calculate the initial evolution of the 
putative Wd1-5 binary (Clark et al. \cite{clark14}).  Unfortunately, to date there is little observational data to favour one 
scenario over the other; indeed it is to be hoped that  
studying binaries within co-eval clusters  may help resolve this issue. 

Focusing   on Wd1, and the evidence appears equivocal. Both the spectroscopic  and evolutionary  mass estimates for Wd1-27 and -30a are sufficiently high (Sect. 4.2 and 4.3) that one must presume a high mass accretion efficiency in a primordial binary in which both components were initially very massive.
Conversely  multiwavelength  observations of the sgB[e] star 
Wd1-9, an object thought to be a binary currently undergoing rapid case A 
mass-transfer, reveals the presence of a cold, dusty circumbinary torus and massive bipolar outflow (Clark et al. 
\cite{clark13}, Fenech et al. \cite{fenech17}). While mass loss through the disc has yet to be quantified, mm-continuum 
and radio recombination line observations suggest a current mass-loss rate of $\sim10^{-4}M_{\odot}$yr$^{-1}$ (Fenech et al. 
\cite{fenech17}) via the bipolar outflow; comparable to the non-conservative  mass-transfer/loss rates of Petrovic et al. (\cite{petrovic}) during certain phases of case A mass-transfer.

Moreover Wd1-27 and -30a are the only clear examples of blue stragglers within Wd1, despite a population of $\gtrsim100$ O9-B4 
stars of luminosity class I-III within Wd1 (Negueruela et al. \cite{iggy10}, Clark et al. in prep.).
This is  despite Schneider et al. 
(\cite{schneider15}) suggesting a peak in their occurrence at the age of Wd1  ($\sim5$Myr), with a ratio of blue 
stragglers to stars below (and within two magnitudes of) the main-sequence turn-off  of $\sim0.2$.
Taken at face value this might be interpreted as potentially disfavouring efficient, quasi-conservative mass-transfer.

However  caution needs to be applied when interpreting such numbers. Firstly,  due to the difficulty of observing cluster 
members in the  crowded core regions, the census of  evolved massive stars within Wd1 is currently incomplete. 
Moreover one could easily  imagine that a subset of putative blue stragglers could have 
already evolved  to the WR phase.  The WR population of Wd1 is known to be binary rich (e.g. Clark et al. 
\cite{clark08}) and at this point (post-)blue stragglers would become  difficult to distinguish from stars evolving via 
alternative channels. Similarly  a fraction of the current OB supergiant population may have already
accreted significant quantities of  mass, even if not appearing as {\em bona fide} blue stragglers at this time\footnote{ 
Unfortunately, the limitations  of our current spectroscopic datasets for many cluster members precludes quantitative determination of  elemental abundances and  rotational velocities, which would help identify candidate post-binary interaction systems.}. Finally the frequency of occurrence of blue 
stragglers is a function of both the mass-transfer efficiency and the properties assumed for the parent binary population.
Schneider et al. (\cite{schneider15}) assume a primordial  binary frequency of 100\%, offering considerable scope for 
reducing the number of massive blue stragglers simply by reducing this fraction; indeed the large number of cluster members
  within Wd1 which appear to have followed a single-star channel reflect this possibility.

Nevertheless, the simple and hopefully uncontroversial result that blue stragglers are present within Wd1 indicates that in 
certain circumstances  mass-transfer from primary to secondary in massive compact binaries can be efficient and lead to 
rejuvenation of the latter. An immediate consequence of this is that, for a subset of objects, massive star 
formation can be  regarded as a two-stage process, with accretion initially forming high-mass progenitor stars and a 
subsequent episode  of binary-driven mass-transfer or merger further increasing their masses. 

Such a pathway also has important implications for the production of post-SNe relativistic remnants. Wd1 serves as an examplar of this process, with the magnetar  CXOU J1647-45 hypothesised to have formed from the core-collapse of a  massive blue straggler progenitor (Clark et al. \cite{clark14}). Indeed, with a mass in the range $\sim51-56.6M_{\odot}$ (evolutionary) to $\sim69^{+37}_{-26}M_{\odot}$ (spectroscopic; Sect. 4.3), Wd1-30a is broadly comparable to predictions for the nature of the companion in  the  putative Wd1-5 binary after rapid, quasi-conservative case-A mass-transfer (Clark et al. \cite{clark14}).  If such a scenario is applicable to Wd1-30a, it will next evolve through an LBV phase at which point a second  interaction  
with the primary will eject  its outer H-rich layers in a common  envelope phase before it proceeds directly to a WR phase and subsequent core-collapse and potential magnetar formation if sufficient mass has been lost from, and angular momentum retained within the core.

Indeed the observational finding that within Wd1 there exist the products of both high- and low-efficiency mass transfer and a single-star channel suggests that in general massive stellar evolution  depends very sensitively on the initial orbital parameters of the binary population and that the concept of a simple one-to-one relation between initial stellar mass and final (relativistic) remnant is likely incorrect.

\subsection{Massive blue stragglers in other stellar aggregates?}

Following the above discussion, are any further examples of massive blue stragglers present in other stellar aggregates? 
An early survey for high-mass blue stragglers was undertaken  by  Mermilliod (\cite{mermilliod}), with Ahumada \& Lapasset 
(\cite{BS}) providing a modern  catalogue of blue stragglers within open clusters\footnote{Schneider et al.
(\cite{schneider15}) report a  deficit of blue stragglers within the young clusters of the Ahumada \& Lapasset (\cite{BS}) census
in comparison to  the results of their simulations.}.  We employ the latter as our primary 
resource, reviewing those clusters with age $\leq10$Myr, corresponding to a main sequence turn-off mass of $\sim20M_{\odot}$. 
The resulting population is supplemented with  the YMC lists of Clark et al. (\cite{clark13}) and the clusters 
and associations 
surveyed by  Massey et al. (\cite{massey95}, \cite{massey00}, \cite{massey01}).

Upon reassessment, the candidacy of many massive blue stragglers is subject to a degree of uncertainty, typically for 
one of three reasons. Firstly the stellar aggregate appears younger than assumed by Ahumada \& 
Lapasset (\cite{BS}) and as a consequence the nature of the blue straggler is uncertain; an example being 
WR 133 (WN5) within NGC6871/Cyg OB3 (Massey et al. 
\cite{massey95}). Secondly the aggregate is potentially non-coeval, such that the putative blue straggler is instead the natural 
product of ongoing star formation.   Cyg OB2 presents a clear example of this, with the presence of O3 supergiants 
reported by Negueruela  et al. (\cite{iggy08}) naturally accommodated under the star formation history subsequently  advocated 
by Wright et al.  (\cite{wright15}). Similar issues afflict NGC6231/Sco OB1, with  studies suggesting a complicated and 
extended star formation history for both cluster and OB association (Clark et al. \cite{clark12} and refs. therein)\footnote{However 
the  authors note that with the exception of HD 152233 (O5.5 III(f)+O7.5)  and HD 152248  (O7.5III(f)+O7 III(f)), NGC6231 
appears  essentialy co-eval, with an age of 5Myr; are  these stars therefore blue stragglers, or are they simply symptomatic 
of an extended episode of star formation?}. Finally  putative blue stragglers may be field stars randomly projected against the 
cluster. 
This is most clearly illustrated by the O6.5 V star HD14434; significantly younger than the $\sim14$Myr age inferred for $\chi$ 
Persei, but considered  a likely interloper (Slesnick et al. \cite{slesnick}, Walborn \cite{walborn}, Currie et al. \cite{currie}).

Nevertheless, the three clusters located within the Galactic centre merit individual discussion, not least because they 
motivated the analysis of Schneider et al. (\cite{schneider14}). These authors suggested that a large number of the most luminous 
stars within the Arches ($9\pm3$) and Quintuplet ($8\pm3$) were likely the rejuvenated products of binary interaction based upon  the 
findings of Martins et al. (\cite{martins08}) and Liermann et al. (\cite{liermann12}). Clark 
et al. (\cite{clark18a}) revisited the Arches, finding that  the treatment of interstellar 
reddening in previous works was oversimplified. This results in significantly underestimated uncertainties in the stellar 
luminosities derived and that, in turn, were used to assert that the population of WNLha stars within the Arches were likely 
binary products. Furthermore Clark et al. (\cite{clark18a}) found a much larger population of mid-O hypergiants than 
Martins et al. (\cite{martins08}) reported, which  appear to  smoothly bridge the divide between O supergiants and WNLha 
stars, implying a close evolutionary relationship. These findings 
in themselves do not rule out that binary products may be present within the Arches, but we find no compelling evidence for 
them at this juncture - and any present would appear to be spectroscopically indistinguishable from other cluster members.

Turning to the Quintuplet and the re-analysis of Clark et al. (\cite{clark18b}) revealed a much more homogeneous stellar population 
than previously reported for the cluster by Liermann et al. (\cite{liermann09}). Critically the observed distribution of 
spectral types closely follows the predictions for the 
evolution of a single $60M_{\odot}$ star (Groh et al. \cite{groh}) up to the H-free WR phase. This includes the Pistol star, 
for which a downwards revision of its luminosity  removes the apparent requirement for 
a binary-modified evolutionary pathway  to explain its presence in the cluster  (Figer et al. \cite{figer}, Najarro et al. \cite{najarro09}). However there is a 
small subset of five stars\footnote{LHO-01 (O7-8 Ia$^+$), -54 
(O7-8Ia$^+$) and -99 (WN8-9ha), qF274 (WN8-9ha) and 406 (O7-8 Ia$^+$).} which, by virtue of their spectral morphologies, appear 
anomalously young (Clark et al. \cite{clark18b}). In the absence of quantitative modeling their origin is uncertain but, as with the cohort of 
WN9-11h/BHGs within Wd1, one might envisage mass-stripping  from the primary in a binary system (Sect. 5.1) since they do not appear significantly hotter than other cluster members (cf. Wd1-27 and -30a).

Next we address the Galactic Centre cluster. As with the Quintuplet and Wd1 it has not proved possible to define the main 
sequence turn-off, but the least evolved objects present appear to be a population of $\sim$O9.5-B2 supergiants, with mean 
$T_{\rm 
eff}\sim27.5$kK and log$(L_{\rm bol}/L_{\odot})\sim5.3$ (Martins et al. \cite{martins07}). However Geballe et al. 
(\cite{geballe}) report a  classification of an O5-6 I-III star associated with the bow-shock IRS8, with modeling 
of IRS8$^*$ suggesting $T_{\rm eff}\sim36{\pm}2$kK and log$(L_{\rm bol}/L_{\odot})\sim5.6{\pm}0.2$\footnote{Following the 
discussions 
regarding extinction towards the Galactic centre in  Clark et al. (\cite{clark18a}), we might anticipate that the errors 
associated with the luminosities derived for these stars are likely to be significantly underestimated.}. The orientation of 
the bow-shock suggests an origin within the Galactic Centre cluster, while the combination of modeling results and  
a cluster age of $6\pm2$Myr is consistent with a blue straggler  identification. However Pfuhl et al. 
(\cite{pfuhl}) note that the relatively large displacement of IRS8$^*$ from the galactic centre and the steep radial profile 
of early-type stars within the central cluster casts some residual doubt on its membership and hence nature. Clearly  radial 
velocity monitoring to identify potential signatures of binarity would be of considerable interest in resolving this issue.

Finally and for completeness we turn to the WN5ha stars within R136, which appear exceptionally massive (Sect. 1). Crowther 
et al.  (\cite{crowther16}) discuss these stars, and the possibility that they are blue stragglers, in depth, suggesting that 
the youth and mass of R136, when combined with  the comparable ages of both the WN5ha and O stars render the possibility 
unlikely.

To summarise: a critical reappraisal of extant data reveals  a comparative lack of confirmed massive blue stragglers
within galactic and Magellanic YMCs - in the sense that stars are {\em unambiguously} hotter and  more luminous and massive that their siblings. Binary interaction may be inferred for a small cohort of Quintuplet members while, if 
physically associated with the Galactic centre cluster, IRS 8$^*$ appears the strongest blue straggler candidate after Wd1-27 
and -30a.
At first glance this finding may appear to be in tension with the predictions of de Mink et al. (\cite{demink}) and Schneider et al. (\cite{schneider14}, \cite{schneider15}). However we highlight that (i) the disparate nature of the data 
employed makes it impossible to systemically evaluate observational biases or uncertainties and (ii) we cannot discount the 
possibility that 
some of  the most massive stars within clusters such as the Arches are indeed the product of  binary mass-transfer or merger, 
 simply that current data do not as yet mandate such an hypothesis. More detailed quantitative analysis, incorporating potential 
 mass transfer diagnostics such as rotational velocity and chemical composition, will be required to confirm this provisional 
conclusion.

\section{Conclusions}

We present a detailed quantitative analysis of multi-wavelength and -epoch data compiled for two members of Wd1 selected on the basis of their anomalously high  X-ray luminosities. Both Wd1-27 and -30a are found to be hotter and more luminous than other cluster members, with 
spectroscopic and evolutionary mass determinations considerably in excess of those inferred for the current supergiant population. These findings imply that both stars are younger than the remaining population of Wd1, which previous studies have suggested is remarkably co-eval (Negueruela et al. \cite{iggy10}, Kudryavtseva et al. \cite{ku}). We conclude that both stars are genuine cluster members from consideration of systemic radial velocities and analysis of both proper motion and parallactic measurements provided by Gaia DR2. This implies that Wd1-27 and -30a are the first examples of (massive) blue stragglers within Wd1. Analysis of both X-ray and  RV data provides no evidence of {\em current} binarity for Wd1-27, although it may not be excluded. Conversely the hard X-ray spectrum of Wd1-30a and the presence of RV variability at over $99$\% likelihood, suggests it is a binary with an orbital  period of $\leq$10days.
Likewise, significant surface nitrogen enhancement in the absence of carbon and oxygen depletion is difficult to understand under single star evolution.

Following the classification and  analysis of Wd-27 and -30a, Wd1 is seen to host a unique population of hypergiants ranging in spectral type from O4-5 Ia$^+$ (Wd1-30a) 
through to F8 Ia$^+$ (Wd1-8a) and, arguably, given the  extreme luminosity of Wd1-26, M6 Ia (Clark et al. \cite{clark10}). 
Such a spread cannot be explained via a single star evolutionary channel for a coeval cohort. Instead we suppose three distinct pathways, with hypergiants of spectral type B5 and later evolving in isolation, the early B hypergiants/WNVLh stars via mass stripping of the 
primary in close binaries  and the O hypergiants considered here by significant mass transfer onto the secondary in compact systems leading to rejuvenation (e.g. Wd1-30a) or, in extreme cases stellar merger (potentially Wd1-27).

The frequency of occurrence of the latter evolutionary channel and the degree of rejuvenation possible is a critical function of how 
much mass massive stars can accrete (and hence how much angular momentum can be  shed and via what mechanism) and the physics 
 of common envelope evolution and stellar merger. 
The extreme current masses suggested for Wd1-27 and -30a imply that mass-transfer must be rather efficient, although  the intense mass-loss  exhibited by the interacting binary Wd1-9 (Clark et al. \cite{clark13}, Fenech et al. \cite{fenech17}) is potentially  in tension with this finding. Likewise Schneider et al. (\cite{schneider15}) suggest that one might expect a large number of rejuvenated binary products within Wd1 at this epoch but we fail to identify any further examples. 
Moreover, with the possible exceptions of the Quintuplet and Galactic centre clusters, there appears to be  a lack of  unambiguous massive blue stragglers in other stellar aggregates; Wd1-27 and -30a appear to represent the most extreme examples of this phenomenon to date. However such a qualitative assessment clearly requires systematic quantitative verification via interrogation  of potential binary-interaction diagnostics such as surface abundances and rotational velocity.

Nevertheless the discovery that such a pathway is viable has important implications, implying that in a subset of cases
massive star formation is a two-stage process, with mass transfer during core-H burning leading to masses significantly in excess 
of the initial `birth-mass' of the recipient. As highlighted in Schneider et al. (\cite{schneider14}) this will impact on the nature of the upper-limit to stellar mass, feedback from such very massive stars and ultimately the nature of their death  (i.e. direct collapse  or pair production SNe) and hence the relative frequency of such events. Regarding the latter and we might expect the massive stars resulting from this evolutionary channel  to be  rapidly rotating with potential consequences for the occurrence of e.g. $\gamma$-ray bursts and the formation of magnetars via the dynamo mechanism.     As a case in point we highlight the apparent similarity of Wd1-30a to 
predictions for the properties of the magnetar progenitor within Wd1 (Clark et al. \cite{clark14}); thus  providing corroboration for the 
formation channel proposed and indicative of the diversity of  physical outcomes rendered possible by binary interaction.

\begin{acknowledgements}
This research was supported by the UK Science and
Technology Facilities Council. and  the the Spanish Government Ministerio de Econom\'{\i}a y Competitivad
(MINECO/FEDER) under grants AYA2015-68012-C2-2-P (Negueruela) and 
ESP2015-65597-C4-1-R and ESP2017-86582-C4-1-R (Najarro).
 This research has made use of the Simbad, Vizier
 and Aladin services developed at the Centre de Donn\'ees Astronomiques de
 Strasbourg, France. This work has made use of data from the European
Space Agency (ESA) mission Gaia (https://www.cosmos.esa.int/gaia), processed by the Gaia
Data Processing and Analysis Consortium (DPAC, https://www.cosmos.esa.int/web/gaia/dpac/consortium). Funding for the 
DPAC has been provided by national institutions, in particular the institutions
participating in the Gaia Multilateral Agreement.

\end{acknowledgements}

{}

\appendix

\section{An analysis of DR2 data on Westerlund~1}

  The second Gaia data release (DR2; Gaia Collaboration et al. \cite{gaia}) has made available precise 
positions, parallaxes and  proper motions for over a billion stars, and thus represents a major step forward in our understanding of stellar 
 physics. Unfortunately, it appears that DR2 measurements in the field of Wd~1 are not necessarily reliable. 
 Extreme examples are the parallaxes to well-known members of the cluster, such as the Wolf-Rayet binary Wd1-241 
(WR77p; $\pi=3.91\pm0.48\:$mas), the yellow hypergiant Wd1-4 ($\pi=0.97\pm0.14\:$mas) and the LBV Wd1-243 
($\pi=0.98\pm0.16\:$mas). 
 These values imply distances that are far too small to be compatible with any previous estimates, but are also impossible to 
 reconcile with the run of extinction with distance observed along the line-of-sight to Wd 1. Specifically Capitanio et al. (\cite{capitanio}) demonstrate that the total extinction out to $\sim1$~kpc in the direction to Wd~1 produces a colour excess of only 
$E(B-V)\sim0.3\pm0.1$, whereas these objects each exhibit $E(B-V)>4$~mag. Moreover, the foreground O9\,Ib supergiant HD~151018, 
which is projected just on top of the cluster, has $E(B-V)\approx0.9$ and  a spectrophotometric distance of 3.1~kpc (M\'{a}iz 
Apell\'{a}niz \& Barb\'{a} \cite{maiz}), suggesting that these cluster members must be at a greater distance still given their reddening; a conclusion incompatible with their parallaxes.

\begin{figure}
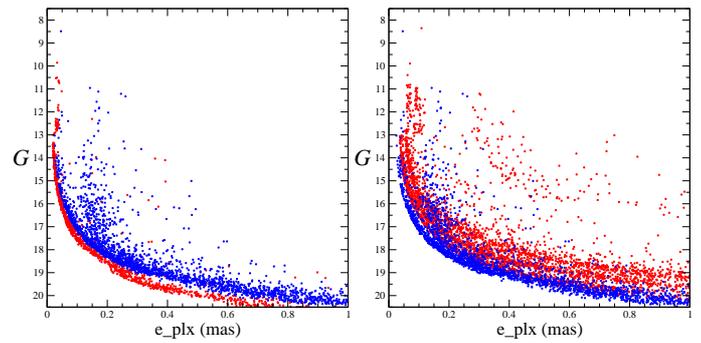

\resizebox{\columnwidth}{!}{\includegraphics[angle=0,clip]{compN7789.eps}\includegraphics[angle=0,clip]{compM11.eps}}
\caption{A comparison of the typical errors in parallax determination within Gaia DR2 for Wd~1 and other 
 crowded regions of the Galactic Plane. \textit{Left panel: }The core of the old open cluster NGC~7789. \textit{Right 
 panel: }The core of the intermediate-age open cluster M11. In both panels, data for Wd~1 is plotted as blue dots, 
 while data for the other cluster is plotted as red dots.}
\label{errors}
\end{figure}

 To explore the source of these unexpected values, we downloaded all the DR2 data within a circle of radius $3\farcm5$ 
 around the nominal centre of Westerlund~1. This area contains more than 3500 Gaia DR2 sources, but only 2265 
 of them have an astrometric solution. If we look at the {\tt astrometric\_excess\_noise\_sig} flag, close to 1300 
 sources -- including almost every source in the central concentration -- have $D>2$, a value suggestive of 
 significant excess noise $\epsilon_{\mathrm{i}}$ for the fit. In fact, less than 400 sources have 
 $\epsilon_{\mathrm{i}}\approx0$, indicating that the residuals of the fit statistically agree with the assumed 
  observational noise. 

The low average quality of the astrometric solutions is made evident in Fig.~\ref{errors}, 
 where  we compare our field to another crowded region of the Galactic Plane, the core of the old open cluster 
 NGC~7789. A circle of the same radius contains more than 1300 sources, of which more than 90\% have astrometric 
 solutions. The left panel of Fig.~\ref{errors} compares the typical errors in parallax at a given magnitude for both 
 fields (plots displaying proper motion errors evidence the same behaviour). Errors are always larger for the Wd~1 
 field and a very significant fraction of stars have much larger errors than the NGC~7789 objects of the same $G$ 
 magnitude. 

Almost all the stars with larger than typical errors lie in the central concentration, suggesting that 
 crowding is the main source of the increased uncertainties. To test this explanation, in the right panel of 
 Fig.~\ref{errors}, we plot the typical errors in an even more crowded region, the central $3\farcm5$ of the 
 intermediate-age cluster M11. This field contains more than 6800 sources, of which nearly 4000 have 
 astrometric solutions. The plot shows not only that errors are typically larger than in Wd~1, but also that several 
"families" of solutions exist at a given $G$ magnitude with different typical errors. It is thus clear that 
 crowding  introduces major uncertainties in DR2 astrometric solutions and therefore individual values must be taken 
 with extreme care.
 
\begin{figure}
\resizebox{\columnwidth}{!}{\includegraphics[angle=0,clip]{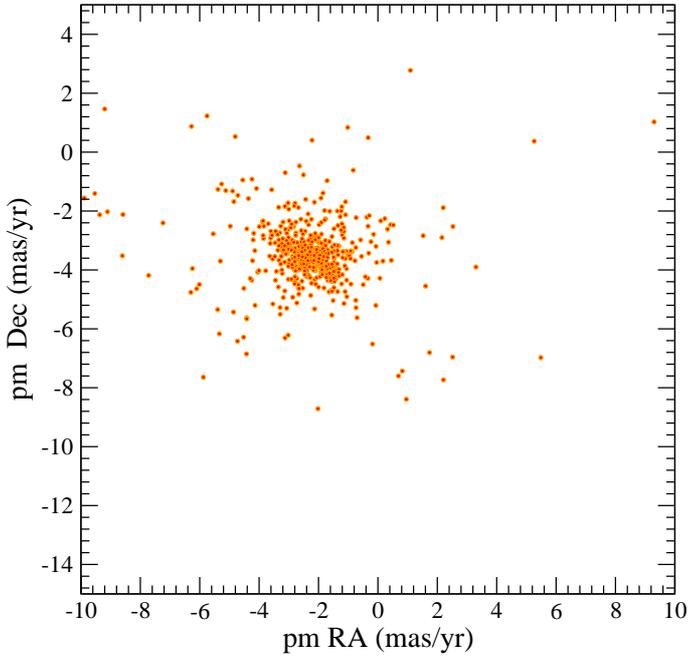}}
\caption{Gaia DR2 proper motions for targets within $3\farcm5$ of
the centre of Wd~1 that display $BP-RP>3.4$, showing the strong concentration of objects around a central value. }
\label{select}
\end{figure}

 Given the huge number of sources and the large uncertainties, in order to identify cluster members, we performed an 
 initial selection by taking only stars with ($B_{\mathrm{P}}-R_{\mathrm{P}})>3.4$, a condition that removes all the 
 foreground population without leaving out any known cluster member (in fact, exploration of DR2 data for a large area 
 around Wd~1 shows that cluster members are the reddest stars with Gaia photometry in the region). When we 
 plotted the resulting sample in the proper motion (pmRA/pmDec) plane, the cluster appears as a strong overdensity 
 (Fig.~\ref{select}), from which we can select candidate members. 

From this initial sample, we calculated the average 
 proper motions by weighting every value with the inverse of its error and then cleaned the sample via an iterative 
 procedure, by discarding outliers and recalculating the average, until the standard deviation of the proper motion 
 values in the sample was comparable to the typical error of an individual value (evaluated as the median of the 
 errors). Removal of the outliers does not imply any judgement on their cluster membership, but was simply intended to 
 define a clean sample of objects with moderately low errors. The procedure proved to be  very robust, with the weighted averages 
 not changing  significantly throughout. We were finally left with a sample of 168 objects, which yielded weighted 
 averages for the proper motions of Wd~1 of   pmRA = $-2.3\:$mas/yr and pmDec =$-3.6\:$mas/yr, with standard deviations of 
  $\sim0.3\:$mas/yr, comparable to the (very large) median errors. For this sample, we calculated a weighted average 
 value of the parallax, finding $\pi=0.19\:$mas. Most of the objects included in the sample have errors comparable to 
 or even larger than their individual parallaxes. We removed those objects that were incompatible with this average within their 
 own uncertainty, coming to a final "clean" sample of 146 objects that are compatible with the average proper motion 
 and parallax values within one sigma. We find  both a weighted and unweighted average parallax of $\pi=0.19\:$mas for this cohort, 
 while the median parallax is $\pi=0.16\:$mas, with a standard deviation of $0.15\:$mas. Taken at face value, these 
 results favour a distance of 5\,--\,6~kpc for Wd~1, in good agreement with estimates based on its high-mass stellar 
 population. However, systematic uncertainties of up to $\pm0.1\:$mas cannot be ruled out (Luri et al. \cite{luri}), implying a 
 range of values (which we will assume are distances not discarded by the astrometry) between 3.5~kpc and unphysically 
 large distances.

 While the average values obtained may be considered reliable since they are based on a large number of objects, any given 
 individual value is suspect, because of the large uncertainties. It is, however, extremely unlikely that an erroneous 
 astrometric solution will result in parameters compatible with cluster averages. For this reason, we can expect 
 objects whose proper motions and parallaxes are consistent with those averages to be cluster members, while a final
 decision on stars with discrepant values must await further Gaia data releases. 

To validate this methodology, we utilised the  photometric
 and spectroscopic censuses of Wd1 presented by   Clark et al. (\cite{clark05}),
 Crowther et al. (\cite{crowther06}), Ritchie et al. (\cite{ritchie09}), and Negueruela et al.  (\cite{iggy10}) to 
produce a master list of `spectroscopic members' - and then inspected their individual DR2 parameters in order to 
determine whether they were flagged as astrometric cluster members. The following stars were found among the "clean" 
sample defined above and  therefore may be considered proper-motion and parallax validated members of 
 the cluster (ordered by brightness): Wd1-57a, -2a, -11, -61a, -52, -56a, -373, -10, -5, -34, -27, -239, -3005, -3004,
 -54, -53, -6a, -74, -3003, -84, -60, -3002, -61b, -58, -17, -1, -59, -63a, -65, -15, -56b, -49, -86, -48, and -228b. We highlight the inclusion of Wd1-27 within this group of stars.

A large subset  of the remaining spectroscopic cluster members  have proper motions that
are compatible (i.e. within twice their individual errors) with the average 
 cluster proper motion. Their parallaxes have very large errors, but again all 
 are within two error bars of the cluster mean; this cohort comprises Wd1-18, -20, -23a, -29, -30a, -37, -38, -43a, 
-50b, -55, -70, -78, and -238 and  WR O, WR R, WR T, and WR V. We emphasise the appearance of Wd1-30a
within this grouping. This subset 
also includes the red   hypergiant Wd1-20 and one of the WR stars furthest from the cluster core (WR T). While the 
yellow  hypergiant Wd1-265 has a proper motion compatible with that of Wd1, as with Wd1-4 it has  an anomalous
 parallax ($\pi=0.80\pm0.17\:$). An additional subset of stars have astrometric solutions
 consistent with the average proper motion of Wd1 at approximately $2\sigma$, although many of them 
are not consonant    with the average cluster parallax; these comprise Wd1-7, -8a, -12a, -13, -21, -19, -24, -33, -35, -39, -43c, 
-46a, -72, -75, and -237 and  WR C   and WR N. This list includes about half the hypergiants and the most distant WR 
star, WR N.

 The list of known members whose DR2 solution is incompatible with the average values comprises Wd1-4, -6b, -8b, -16a, -26, 
 -28, -31a, -32, -42a, -46b, -50a, -62a, -71, -238, -241, and  -243 and  WR B, WR D, WR G, WR M, and WR P. Finally, a small number of stars have no astrometric solution in DR2, namely:  Wd1-9, -14a, -41, -43b, and -44 and  WR H, WR I, WR J, and WR K.

\begin{figure}
\includegraphics[width=8cm,angle=0]{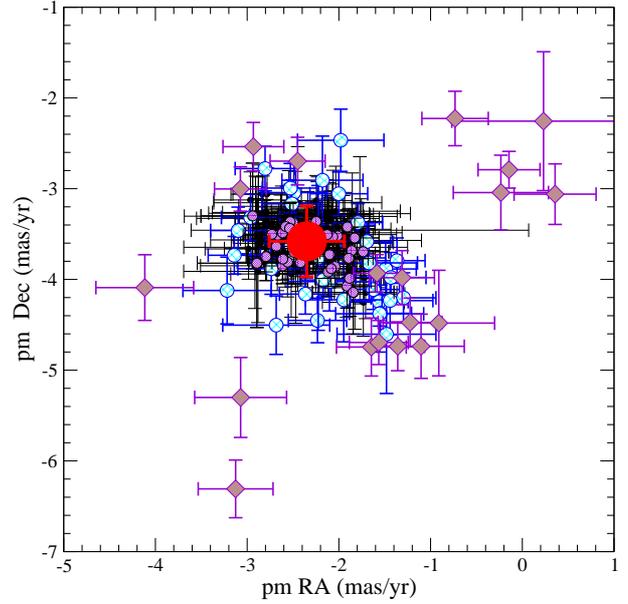}
   \caption{Gaia DR2 proper motions for possible members of Wd~1.  The small filled (violet) circles are 
spectroscopically selected cluster members that form a subset of the "clean" population defined on the basis of proper motions and 
parallaxes; the remaining members of this grouping (i.e. those stars without spectral classifications) are shown as black dots.
Larger (blue) circles are spectroscopically selected members with astrometric properties compatible with the cluster averages within approximately two  of their own error bars. Diamonds are spectroscopic cluster 
members that are not compatible at 2\,$\sigma$ in at least one of  the three astrometric values. The large (red) 
circle represents the average cluster value. Its error bars are the median  value of the errors for the sample of 
objects used to calculate it.
           }
   \label{pms}
    \end{figure}

 Figure~\ref{pms} shows the proper motion plane for stars that have been classified as members. The vast 
 majority are compatible within two of their error bars with the average cluster proper motions; indeed re-evaluating the cluster proper
motion including {\em all} the spectroscopic members yields proper motions of pmRA = $-2.1\:$mas/yr and pmDec =$-3.7\:$mas/yr, which remain compatible with those derived from the "clean" sample within the respective errors.   Agreement with the  average parallax is worse, but even including objects with unphysically 
high parallaxes, such as the  aforementioned Wd1-4, -241, and -243, only shifts the weighted average from $0.19$ to $0.22$~mas, 
All the astrometric and photometric values available within DR2 for known cluster members are listed in Table~\ref{dr2}.

Finally, we highlight that  none of the very luminous  blue, yellow or red hypergiants within Wd1 appear in the list
of proper-motion and parallax validated cluster members, the brightest star being the B4\,Ia supergiant Wd1-57a. 
 Nevertheless, aside from their high reddenings, 
 there are compelling observational reasons to believe that they are indeed cluster members. Firstly the Gaia 
DR2 includes radial velocities for four of the cool hypergiants (Table~\ref{dr2}) which are all
 consistent with the cluster average and velocity dispersion (Ritchie et al. \cite{ritchie09}, Clark et al. \cite{clark14}, in 
prep.). Secondly, their locations in  
both  colour/magnitude and HR diagrams are consonant with the proper-motion validated cohort (Clark et al. \cite{clark05}, 
Negueruela et al. \cite{iggy10}, Ritchie et al. \cite{ritchie10}). There is also a  smooth and continuous 
progression in spectroscopic morphologies from the early- and mid-B supergiants within Wd1 through to the late B hypergiants, yellow hypergiants and finally red supergiants (Clark et al. \cite{clark05}, Negueruela et al. \cite{iggy10}). Finally 
mm- and radio-continuum observations  reveal prominent cometary nebulae associated with the majority of the cool hypergiants
which are all orientated towards the cluster core; a phenomenon attributed to the ionisation and sculpting of their stellar outflows by the radiation and wind pressure of the  host of WR and O-type stars within Wd1 (Andrews et al. \cite{andrews}, Fenech et al. \cite{fenech18}). This behaviour, replicating that of  the red supergiant GC IRS7 (associated with the nuclear star cluster; 
Yusef-Zadeh \& Morris \cite{YZ}), conclusively proves a physical  association between cluster and the  hypergiants.

 To summarise, our analysis shows that both Wd1-27 and -30a are confirmed as proper-motion and parallax validated 
cluster members. The former is part of 
the  core "clean" sample used to define the averages, while the second is compatible within approximately 2\,$\sigma$. At 
 present, we can only speculate about the reasons why the majority of the most luminous spectroscopic cluster 
members have untenably high parallaxes, although we highlight that many of  these stars  are sufficiently large that
their disks may be  resolved by  Gaia, even at a distance of $\sim5$~kpc. Moreover  many of the wider cluster population 
are found to be binaries (cf. Clark et al. \cite{clark08}, Ritchie et al. \cite{ritchie09}), an occurence which is known to affect the reliability of Gaia DR2 astrometric solutions. We suspect  
 these circumstances conspire with the very high stellar density to yield unreliable astrometric solutions. Future 
 Gaia data releases will undoubtedly provide a much sharper view of membership in Wd~1, while very likely 
 confirming the average cluster values determined here.

\longtab{1}{
\begin{longtable}[h]{l c c c | c c  c c | c}
\caption{Gaia DR2 data for previously catalogued members of Westerlund~1.\label{dr2}}\\
\hline\hline
Name	&Plx (mas)	 &pmRA (mas)		&pmDE (mas)		&$G$ mag	   &BP mag	      &RP mag	         &BP$-$RP&RV (km\,s$^{-1}$)\\
\hline
\endfirsthead
\caption{continued.}\\
\hline
\hline
Name	&Plx (mas)	 &pmRA (mas)		&pmDE (mas)		&$G$ mag	   &BP mag	      &RP mag	         &BP$-$RP&RV (km\,s$^{-1}$)\\
\hline
\endhead
\hline
\endfoot
 Wd1-1	&0.10$\pm$0.16   &$-$1.88$\pm$0.37   	&$-$3.99$\pm$0.42	&15.597$\pm$0.001  &18.427$\pm$0.016  
&14.077$\pm$0.004  &4.35   &	\\
Wd1-2a	&0.19$\pm$0.12   &$-$1.74$\pm$0.28   	&$-$3.70$\pm$0.22	&13.709$\pm$0.002  &16.641$\pm$0.021  
&12.179$\pm$0.008  &4.46   &	\\
Wd1-4	&0.97$\pm$0.14   &$$-$$2.45$\pm$0.30    &$-$2.69$\pm$0.26	&10.955$\pm$0.002  &14.463$\pm$0.012  &9.528 
$\pm$0.012  &4.93   &$-$47.4$\pm$0.4\\
Wd1-5	&0.24$\pm$0.14   &$-$2.40$\pm$0.28   	&$-$3.56$\pm$0.21	&14.464$\pm$0.002  &17.553$\pm$0.016  
&12.913$\pm$0.010  &4.64   &	\\
Wd1-6a	&0.10$\pm$0.16   &$-$2.42$\pm$0.31   	&$-$3.82$\pm$0.23	&15.198$\pm$0.004  &18.374$\pm$0.026  
&13.555$\pm$0.014  &4.82   &	\\
Wd1-6b	&$-$0.09$\pm$0.28&$$-$$4.12$\pm$0.53  	&$-$4.09$\pm$0.36	&17.297$\pm$0.006  &	  	      &      		 
&	 &	 \\
Wd1-7	&0.48$\pm$0.16   &$-$1.30$\pm$0.35    	&$-$4.20$\pm$0.28	&11.952$\pm$0.003  &15.524$\pm$0.014  
&10.337$\pm$0.010  &5.19   &	 \\
Wd1-8a	&0.72$\pm$0.17   &$-$1.56$\pm$0.32    	&$-$4.25$\pm$0.27	&11.816$\pm$0.001  &15.434$\pm$0.015  
&10.211$\pm$0.008  &5.22   &$-$43.6$\pm$3.4\\
Wd1-8b	&$-$0.05$\pm$0.12&$-$0.15$\pm$0.34  	&$-$2.79$\pm$0.20	&13.447$\pm$0.003  &16.470$\pm$0.018  
&11.808$\pm$0.013  &4.66   &	 \\
Wd1-9	&	     	 &                      &     	 		&13.522$\pm$0.002  &16.769$\pm$0.018  
&11.844$\pm$0.013  &4.93   &	 \\
Wd1-10	&0.27$\pm$0.15   &$-$1.83$\pm$0.27   	&$-$3.54$\pm$0.22	&14.400$\pm$0.001  &18.009$\pm$0.016  
&12.790$\pm$0.007  &5.22   &	\\
Wd1-11	&0.16$\pm$0.15   &$-$2.67$\pm$0.38   	&$-$3.51$\pm$0.22	&13.870$\pm$0.002  &17.153$\pm$0.026  
&12.308$\pm$0.010  &4.85   &	\\
Wd1-12a	&1.06$\pm$0.17   &$-$1.49$\pm$0.33    	&$-$3.89$\pm$0.29	&12.379$\pm$0.002  &16.596$\pm$0.019  
&10.617$\pm$0.016  &5.98   &	 \\
Wd1-13	&0.18$\pm$0.16   &$-$2.23$\pm$0.32    	&$-$4.45$\pm$0.25	&14.038$\pm$0.004  &17.199$\pm$0.016  
&12.472$\pm$0.013  &4.73   &	 \\
Wd1-14a	&	     	 &                 	&     	 		&16.402$\pm$0.004  &	  	      &     		 
&	 &	 \\
Wd1-15	&0.10$\pm$0.18   &$-$1.84$\pm$0.45   	&$-$4.14$\pm$0.46	&15.776$\pm$0.001  &18.736$\pm$0.046  
&14.155$\pm$0.007  &4.58   &	\\
Wd1-16a	&0.58$\pm$0.17   &$-$1.65$\pm$0.38   	&$-$4.75$\pm$0.32	&11.847$\pm$0.002  &15.780$\pm$0.025  
&10.235$\pm$0.012  &5.54   &	 \\
Wd1-17	&$-$0.01$\pm$0.15&$-$2.08$\pm$0.25   	&$-$3.88$\pm$0.20	&15.583$\pm$0.001  &18.823$\pm$0.030  
&13.957$\pm$0.009  &4.87   &	\\
Wd1-18	&0.49$\pm$0.20   &$-$2.96$\pm$0.32   	&$-$3.32$\pm$0.26	&14.283$\pm$0.001  &17.288$\pm$0.023  
&12.708$\pm$0.008  &4.58   &	 \\
Wd1-19	&$-$0.34$\pm$0.22&$-$1.43$\pm$0.38   	&$-$3.98$\pm$0.28	&14.391$\pm$0.002  &18.026$\pm$0.022  
&12.679$\pm$0.015  &5.35   &	  \\
Wd1-20	&0.26$\pm$0.26   &$-$1.67$\pm$0.53   	&$-$3.82$\pm$0.42	&13.592$\pm$0.001  &19.991$\pm$0.125  
&11.743$\pm$0.016  &8.25   &	     \\
Wd1-21	&0.45$\pm$0.12   &$-$3.14$\pm$0.29   	&$-$3.73$\pm$0.21	&14.813$\pm$0.001  &18.174$\pm$0.039  
&13.204$\pm$0.008  &4.97   &	     \\
Wd1-23a	&0.01$\pm$0.12   &$-$1.79$\pm$0.27   	&$-$3.37$\pm$0.21	&14.040$\pm$0.001  &17.699$\pm$0.014  
&12.404$\pm$0.012  &5.30   &	     \\
Wd1-24	&0.11$\pm$0.13   &$-$1.63$\pm$0.29   	&$-$4.11$\pm$0.22	&15.261$\pm$0.001  &18.660$\pm$0.028  
&13.643$\pm$0.007  &5.02   &	     \\
Wd1-26	&0.68$\pm$0.25   &$-$0.91$\pm$0.61   	&$-$4.48$\pm$0.58	&11.211$\pm$0.006  &16.379$\pm$0.058  &9.649 
$\pm$0.010  &6.73   &$-$49.4$\pm$1.9\\
Wd1-27	&0.16$\pm$0.14   &$-$2.39$\pm$0.26   	&$-$3.72$\pm$0.20	&14.766$\pm$0.001  &17.895$\pm$0.028  
&13.158$\pm$0.012  &4.74   &	\\
Wd1-28	&$-$0.42$\pm$0.31&1.74 $\pm$0.73   	&$-$6.81$\pm$0.81	&13.615$\pm$0.003  &16.828$\pm$0.015  
&12.008$\pm$0.006  &4.82   &	\\
Wd1-29	&0.37$\pm$0.19   &$-$2.18$\pm$0.44   	&$-$2.91$\pm$0.49	&15.408$\pm$0.001  &18.602$\pm$0.075  
&13.713$\pm$0.014  &4.89   &	\\
Wd1-30a	&0.06$\pm$0.16   &$-$2.37$\pm$0.28   	&$-$4.16$\pm$0.22	&15.165$\pm$0.001  &18.404$\pm$0.021  
&13.532$\pm$0.007  &4.87   &	\\
Wd1-31a	&0.37$\pm$0.12   &$-$1.59$\pm$0.26   	&$-$3.93$\pm$0.21	&14.741$\pm$0.002  &17.811$\pm$0.094  
&12.915$\pm$0.032  &4.90   &	\\
Wd1-32	&1.23$\pm$0.17   &$-$1.31$\pm$0.35   	&$-$3.98$\pm$0.30	&11.115$\pm$0.002  &15.197$\pm$0.014  &9.616 
$\pm$0.010  &5.58   &	\\
Wd1-33	&0.62$\pm$0.20   &$-$2.68$\pm$0.58   	&$-$4.50$\pm$0.32	&12.032$\pm$0.002  &15.615$\pm$0.015  
&10.436$\pm$0.007  &5.18   &	\\
Wd1-34	&0.06$\pm$0.19   &$-$2.94$\pm$0.56   	&$-$3.30$\pm$0.58	&14.736$\pm$0.001  &18.043$\pm$0.057  
&13.069$\pm$0.019  &4.97   &	\\
Wd1-35	&$-$0.14$\pm$0.29&$-$1.48$\pm$0.54   	&$-$4.60$\pm$0.65	&15.305$\pm$0.001  &18.712$\pm$0.022  
&13.664$\pm$0.007  &5.05   &	\\
Wd1-37	&$-$0.70$\pm$0.28&$-$2.10$\pm$0.45   	&$-$4.02$\pm$0.32	&15.683$\pm$0.004  &18.827$\pm$0.042  
&13.896$\pm$0.017  &4.93   &	\\
Wd1-38	&$-$0.18$\pm$0.17&$-$1.95$\pm$0.42   	&$-$4.23$\pm$0.46	&15.734$\pm$0.001  &18.956$\pm$0.041  
&13.996$\pm$0.023  &4.96   &	\\
Wd1-39	&0.33$\pm$0.14   &$-$3.10$\pm$0.30   	&$-$3.46$\pm$0.26	&15.977$\pm$0.001  &19.088$\pm$0.051  
&14.233$\pm$0.018  &4.85   &	\\
Wd1-41	&	       	 &         	     	&	  		&14.746$\pm$0.004  &17.785$\pm$0.018  
&13.056$\pm$0.007  &4.73   &	    \\
Wd1-42a	&$-$0.15$\pm$0.17&0.36 $\pm$0.45   	&$-$3.06$\pm$0.33	&12.827$\pm$0.003  &17.095$\pm$0.015  
&11.131$\pm$0.013  &5.96   &	\\
Wd1-43a	&$-$0.45$\pm$0.17&$-$2.74$\pm$0.36   	&$-$3.88$\pm$0.28	&14.501$\pm$0.003  &17.822$\pm$0.058  
&12.449$\pm$0.032  &5.37   &	\\
Wd1-43b	&	       	 &         		&	  		&14.745$\pm$0.003  &	  	      &	  		 
&       &        \\
Wd1-43c	&$-$0.19$\pm$0.21&$-$1.98$\pm$0.47   	&$-$2.47$\pm$0.34	&15.587$\pm$0.003  &18.299$\pm$0.023  
&13.890$\pm$0.011  &4.41   &	\\
Wd1-44	&	       	 &       		&        	   	&14.557$\pm$0.008  &18.555$\pm$0.033  
&12.834$\pm$0.023  &5.72   &	   \\
Wd1-46a	&0.41$\pm$0.18   &$-$1.37$\pm$0.31   	&$-$3.79$\pm$0.25	&14.562$\pm$0.001  &18.474$\pm$0.035  
&12.910$\pm$0.008  &5.56   &	\\
Wd1-46b	&1.65$\pm$0.26   &$-$0.23$\pm$0.52   	&$-$3.04$\pm$0.41	&15.923$\pm$0.003  &18.161$\pm$0.099  
&14.176$\pm$0.013  &3.98   &	\\
Wd1-48	&0.08$\pm$0.14   &$-$2.56$\pm$0.29   	&$-$3.81$\pm$0.23	&15.979$\pm$0.001  &19.622$\pm$0.057  
&14.334$\pm$0.008  &5.29   &	\\
Wd1-49	&0.00$\pm$0.18   &$-$2.49$\pm$0.36   	&$-$3.55$\pm$0.26	&15.808$\pm$0.001  &18.696$\pm$0.033  
&14.148$\pm$0.014  &4.55   &	\\
Wd1-50a	&0.26$\pm$0.12   &$-$3.07$\pm$0.28   	&$-$3.00$\pm$0.24	&15.730$\pm$0.001  &18.681$\pm$0.037  
&14.093$\pm$0.009  &4.59   &	\\
Wd1-50b	&0.59$\pm$0.14   &$-$2.47$\pm$0.33   	&$-$3.56$\pm$0.25	&16.620$\pm$0.004  &	 	      & 		 
&	 &	\\
Wd1-52	&0.39$\pm$0.12   &$-$2.02$\pm$0.25   	&$-$3.52$\pm$0.20	&13.979$\pm$0.002  &17.438$\pm$0.018  
&12.386$\pm$0.009  &5.05   &	\\
Wd1-53	&0.29$\pm$0.14   &$-$2.50$\pm$0.28   	&$-$3.37$\pm$0.23	&15.172$\pm$0.002  &18.559$\pm$0.036  
&13.596$\pm$0.010  &4.96   &	\\
Wd1-54	&0.21$\pm$0.15   &$-$1.86$\pm$0.31   	&$-$3.68$\pm$0.25	&15.111$\pm$0.001  &18.731$\pm$0.034  
&13.494$\pm$0.007  &5.24   &	\\
Wd1-55	&0.09$\pm$0.12   &$-$2.17$\pm$0.25   	&$-$4.01$\pm$0.19	&14.671$\pm$0.001  &17.739$\pm$0.030  
&13.142$\pm$0.007  &4.60   &	\\
Wd1-56a	&0.24$\pm$0.15   &$-$1.91$\pm$0.31   	&$-$4.07$\pm$0.24	&14.088$\pm$0.001  &17.353$\pm$0.025  
&12.518$\pm$0.007  &4.84   &	\\
Wd1-56b	&0.33$\pm$0.10   &$-$2.61$\pm$0.22   	&$-$3.49$\pm$0.17	&15.779$\pm$0.001  &18.878$\pm$0.064  
&14.204$\pm$0.008  &4.67   &	\\
Wd1-57a	&0.32$\pm$0.16   &$-$2.10$\pm$0.32   	&$-$3.73$\pm$0.26	&13.047$\pm$0.002  &16.251$\pm$0.033  
&11.466$\pm$0.007  &4.79   &	\\
Wd1-58	&0.19$\pm$0.13   &$-$1.89$\pm$0.27   	&$-$3.77$\pm$0.21	&15.570$\pm$0.001  &18.894$\pm$0.037  
&13.960$\pm$0.006  &4.93   &	\\
Wd1-59	&0.13$\pm$0.13   &$-$2.53$\pm$0.27   	&$-$3.36$\pm$0.21	&15.631$\pm$0.001  &18.773$\pm$0.031  
&13.983$\pm$0.008  &4.79   &	\\
Wd1-60	&0.10$\pm$0.11   &$-$2.04$\pm$0.24   	&$-$3.86$\pm$0.19	&15.306$\pm$0.001  &18.581$\pm$0.015  
&13.734$\pm$0.006  &4.85   &	\\
Wd1-61a	&0.31$\pm$0.11   &$-$1.90$\pm$0.22   	&$-$3.77$\pm$0.17	&13.967$\pm$0.002  &17.151$\pm$0.019  
&12.396$\pm$0.010  &4.75   &	\\
Wd1-61b	&0.25$\pm$0.14   &$-$2.07$\pm$0.30   	&$-$3.53$\pm$0.24	&15.337$\pm$0.002  &18.608$\pm$0.034  
&13.721$\pm$0.010  &4.89   &	\\
Wd1-62a	&0.62$\pm$0.35   &9.30 $\pm$0.72   	&1.03 $\pm$0.55		&15.675$\pm$0.005  &18.769$\pm$0.047  
&13.895$\pm$0.018  &4.87   &	\\
Wd1-63a	&0.03$\pm$0.11   &$-$1.91$\pm$0.23   	&$-$3.42$\pm$0.19	&15.680$\pm$0.001  &18.358$\pm$0.039  
&14.071$\pm$0.022  &4.29   &	\\
Wd1-65	&0.25$\pm$0.10   &$-$2.79$\pm$0.22   	&$-$3.75$\pm$0.17	&15.700$\pm$0.001  &18.829$\pm$0.025  
&14.135$\pm$0.007  &4.69   &	 \\
Wd1-70	&0.43$\pm$0.15   &$-$2.53$\pm$0.31   	&$-$2.99$\pm$0.27	&13.350$\pm$0.002  &16.921$\pm$0.016  
&11.753$\pm$0.010  &5.17   &	\\
Wd1-71	&$-$0.09$\pm$0.14&$-$1.36$\pm$0.29   	&$-$4.74$\pm$0.27	&13.246$\pm$0.001  &17.009$\pm$0.018  
&11.634$\pm$0.008  &5.37   &	\\
Wd1-72 	&0.07$\pm$0.16   &$-$1.55$\pm$0.34   	&$-$4.37$\pm$0.31	&15.837$\pm$0.002  &19.542$\pm$0.055  
&14.107$\pm$0.010  &5.44   &	\\
Wd1-74	&0.06$\pm$0.20   &$-$2.89$\pm$0.65   	&$-$3.82$\pm$0.71	&15.205$\pm$0.001  &18.332$\pm$0.019  
&13.639$\pm$0.007  &4.69   &	\\
Wd1-75	&0.14$\pm$0.22   &$-$1.44$\pm$0.47   	&$-$4.23$\pm$0.38	&14.702$\pm$0.002  &20.910$\pm$0.144  
&12.851$\pm$0.016  &8.06   &$-$50.6$\pm$2.9\\
Wd1-78	&0.32$\pm$0.12   &$-$1.70$\pm$0.28   	&$-$3.58$\pm$0.21	&13.954$\pm$0.002  &17.101$\pm$0.012  
&12.412$\pm$0.007  &4.69   &	\\
Wd1-84	&0.25$\pm$0.09   &$-$2.36$\pm$0.20   	&$-$3.48$\pm$0.16	&15.298$\pm$0.001  &17.950$\pm$0.011  
&13.816$\pm$0.005  &4.13   &	\\
Wd1-86	&0.29$\pm$0.11   &$-$2.58$\pm$0.24   	&$-$3.42$\pm$0.18	&15.916$\pm$0.001  &18.833$\pm$0.026  
&14.389$\pm$0.006  &4.44   &	\\
Wd1-228b	&0.19$\pm$0.13   &$-$2.32$\pm$0.26   	&$-$3.74$\pm$0.20	&16.100$\pm$0.001  &18.596$\pm$0.081  
&14.577$\pm$0.011  &4.02   &	\\
Wd1-238	&0.50$\pm$0.14   &$-$2.49$\pm$0.25   	&$-$3.75$\pm$0.18	&14.825$\pm$0.001  &17.586$\pm$0.015  
&13.325$\pm$0.009  &4.26   &	\\
Wd1-237	&1.64$\pm$0.26   &$-$1.56$\pm$0.48   	&$-$4.38$\pm$0.42	&11.325$\pm$0.009  &16.857$\pm$0.055  
&9.661 $\pm$0.016  &7.20   &	\\
Wd1-238	&$-$0.29$\pm$0.19&$-$3.13$\pm$0.41   	&$-$6.31$\pm$0.32	&14.454$\pm$0.001  &17.495$\pm$0.019  
&12.904$\pm$0.005  &4.59   &	\\
Wd1-239	&0.09$\pm$0.14   &$-$2.49$\pm$0.28   	&$-$3.70$\pm$0.23	&14.827$\pm$0.001  &17.899$\pm$0.025  
&13.257$\pm$0.007  &4.64   &	\\
Wd1-241	&3.91$\pm$0.48   &0.23 $\pm$1.00   	&$-$2.26$\pm$0.77	&15.008$\pm$0.006  &18.101$\pm$0.015  
&13.368$\pm$0.005  &4.73   &	\\
Wd1-243	&0.98$\pm$0.16   &$-$0.73$\pm$0.36   	&$-$2.23$\pm$0.30	&11.558$\pm$0.003  &15.271$\pm$0.015  &9.983 
$\pm$0.015  &5.29   &	\\
Wd1-265	&0.80$\pm$0.17   &$-$2.49$\pm$0.35   	&$-$3.04$\pm$0.32	&12.502$\pm$0.001  &16.957$\pm$0.015  
&10.808$\pm$0.016  &6.15   &$-$41.3$\pm$1.6\\
Wd1-373	&0.28$\pm$0.10   &$-$2.61$\pm$0.19   	&$-$3.85$\pm$0.16	&14.221$\pm$0.002  &17.107$\pm$0.015  
&12.712$\pm$0.007  &4.40   &	\\
Wd1-3002	&0.18$\pm$0.14   &$-$2.73$\pm$0.29   	&$-$3.50$\pm$0.23	&15.321$\pm$0.001  &18.876$\pm$0.026  
&13.724$\pm$0.007  &5.15   &	\\
Wd1-3003	&0.29$\pm$0.16   &$-$2.12$\pm$0.36   	&$-$3.92$\pm$0.27	&15.284$\pm$0.002  &18.989$\pm$0.023  
&13.666$\pm$0.008  &5.32   &	\\
Wd1-3004	&0.16$\pm$0.13   &$-$2.25$\pm$0.27   	&$-$3.36$\pm$0.21	&14.973$\pm$0.002  &18.808$\pm$0.025  
&13.349$\pm$0.011  &5.46   &	\\
Wd1-3005	&0.28$\pm$0.11   &$-$2.68$\pm$0.22   	&$-$3.64$\pm$0.18	&14.855$\pm$0.001  &17.752$\pm$0.016  
&13.346$\pm$0.008  &4.41   &	\\
\noalign{\smallskip}
WR B	&$-$0.09$\pm$0.15&$-$1.57$\pm$0.32   	&$-$4.70$\pm$0.24	&16.483$\pm$0.004  &20.503$\pm$0.118  &14.754$\pm$0.017  &5.75   &	\\
WR C	&$-$0.03$\pm$0.22&$-$3.22$\pm$0.48   	&$-$4.12$\pm$0.37	&16.802$\pm$0.003  &19.883$\pm$0.112  &14.709$\pm$0.026  &5.17   &	\\
WR D	&$-$0.61$\pm$0.26&$-$3.07$\pm$0.50   	&$-$5.30$\pm$0.44	&17.528$\pm$0.004  &19.945$\pm$0.130  &15.693$\pm$0.015  &4.25   &	\\
WR G	&$-$0.55$\pm$0.22&$-$1.10$\pm$0.47   	&$-$4.74$\pm$0.35	&16.746$\pm$0.003  &19.842$\pm$0.082  &14.959$\pm$0.012  &4.88   &	\\
WR H	&		 &       		&      			&16.224$\pm$0.009  &19.977$\pm$0.125  &14.271$\pm$0.009  &5.71   &	\\
WR I	&	      	 &       		&         		&18.263$\pm$0.019  &20.904$\pm$0.103  &16.075$\pm$0.018  &4.83   &	\\
WR J	&	      	 &       		&         		&16.955$\pm$0.012  &	 	      &      		 &	 &	\\
WR K	&	      	 &       		&         		&16.801$\pm$0.004  &	  	      &     		 &	 &	\\
WR M	&0.29$\pm$0.16   &$-$2.93$\pm$0.33   	&$-$2.54$\pm$0.27	&15.979$\pm$0.002  &19.683$\pm$0.072  &14.306$\pm$0.015  &5.38   &	\\
WR N	&$-$0.28$\pm$0.18&$-$2.81$\pm$0.33	&$-$2.78$\pm$0.25	&15.230$\pm$0.003  &	  	      &     		 &	 &	\\
WR O	&0.20$\pm$0.15   &$-$2.00$\pm$0.32   	&$-$3.06$\pm$0.24	&16.073$\pm$0.002  &19.522$\pm$0.040  &14.453$\pm$0.009  &5.07   &	\\
WR P	&0.09$\pm$0.16   &$-$1.22$\pm$0.35   	&$-$4.47$\pm$0.27	&16.231$\pm$0.003  &19.282$\pm$0.038  &14.396$\pm$0.023  &4.89   &	\\
WR R	&0.16$\pm$0.27   &$-$2.23$\pm$0.73   	&$-$3.76$\pm$0.80	&16.941$\pm$0.003  &	  	      &    		 &	 &	\\
WR T	&0.24$\pm$0.11   &$-$2.52$\pm$0.23   	&$-$3.63$\pm$0.17	&14.976$\pm$0.001  &17.968$\pm$0.012  &13.447$\pm$0.007  &4.52   &	\\
WR V	&0.53$\pm$0.16   &$-$2.51$\pm$0.33   	&$-$3.17$\pm$0.26	&15.802$\pm$0.004  &	  	      &     		 &	 &	\\
WR U	&0.44$\pm$0.16   &$-$2.54$\pm$0.41   	&$-$3.44$\pm$0.27	&16.259$\pm$0.004  &19.205$\pm$0.131  &14.330$\pm$0.054  &4.88   &	 \\
\noalign{\smallskip}
\hline
\end{longtable}}
								     
\section{Model results for Wd1-30a via fitting the Br$\gamma$ line}

\begin{figure*}
\includegraphics[width=14cm,angle=0]{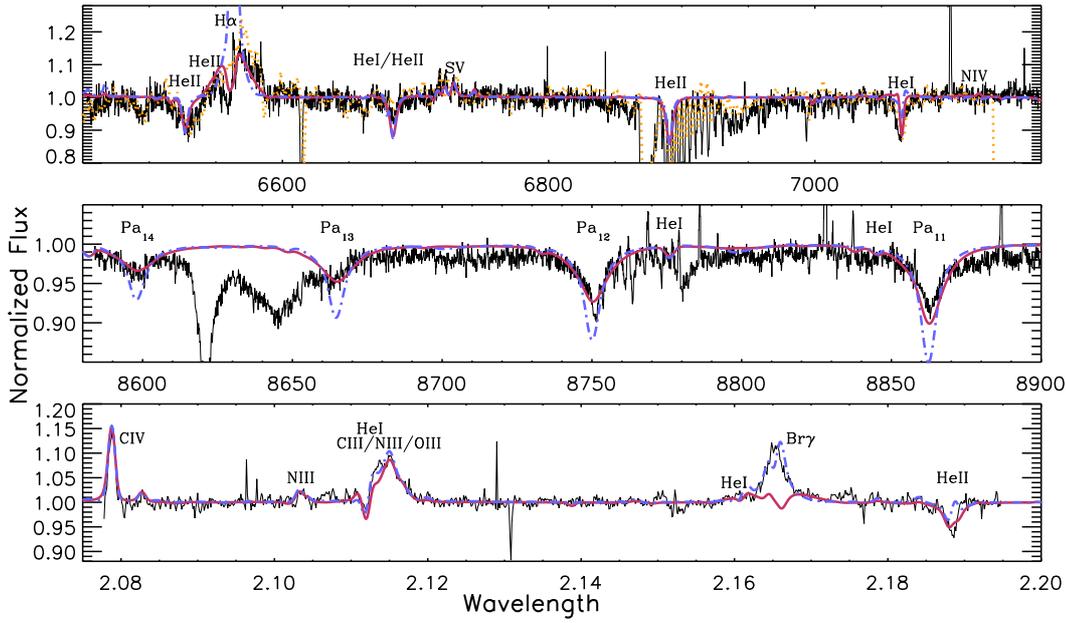}
\caption{Comparison of synthetic spectra of Wd1-30a derived from the best-fit model-atmosphere solution (red line)
 and the  alternative optimised to fit the K-band spectrum and in paticular the Br$\gamma$ profile (blue line; see Sect.
4.3 for details).  Observational data are presented in black, with an additional R-band spectrum (orange) overplotted to demonstrate the variability in the H$\alpha$ profile (spectra from 2004 June 12 \& 13). While it is possible to reproduce the Br$\gamma$ emission it comes at the cost of greatly overestimating the strength of H$\alpha$ emission and the depth of the photospheric  Paschen series lines. For completeness we note that such a model implies a cooler ($T_{\rm eff}\sim34$kK versus 37.25kK) and  lower luminosity 
(log$(L_{\rm bol}/L_{\odot})\sim5.7$ versus 5.89) star with a slower wind ($V_{\infty}\sim800$kms$^{-1}$ versus 1200kms$^{-1}$).
The units of wavelength for the top and middle panels are Angstroms  
and the bottom panel microns.}
\end{figure*}

\end{document}